\documentclass[aps,amssymb,amsmath,twocolumn]{revtex4}

\usepackage[latin1] {inputenc}
\usepackage{bm}
\usepackage[dvips]{graphicx}

\begin{document}

\title{Spontaneous creation of macroscopic flow and metachronal waves in an array of cilia }

\author{Boris Guirao}
\affiliation{Laboratoire Physico-chimie Curie (UMR 168), Institut
Curie, 26 rue d'Ulm, 75248 Paris Cedex 05, France}

\author{Jean-Fran\c cois Joanny}
\affiliation{Laboratoire Physico-chimie Curie (UMR 168), Institut
Curie, 26 rue d'Ulm, 75248 Paris Cedex 05, France}

\date{\today }

\begin{abstract}
Cells or bacteria carrying cilia on their surface show many
striking features : alignment of cilia in an array, two-phase
asymmetric beating for each cilium, coordination between cilia and
existence of metachronal waves with a constant phase difference
between two adjacent cilia. We give simple theoretical arguments
based on hydrodynamic coupling and an internal mechanism of the
cilium derived from the behavior of a collection of molecular
motors, to account qualitatively for these cooperative features.
Hydrodynamic interactions can lead to the alignment of an array of
cilia. We study the effect of a transverse external flow  and
obtain a two-phase asymmetrical beating, faster along the flow and
slower against the flow, proceeding around an average curved
position. We show that an aligned array of cilia is able to
spontaneously break the left-right symmetry and to create a global
average flow. Metachronism arises as a local minimum of the
beating threshold and leads to a rather constant flow.
\end{abstract}

\maketitle


\section{Introduction}

Many cells and bacteria have cilia or flagella on their surfaces.
Examples are sperm cells which have one flagellum used for
propulsion, the green alga \textit{Chlamydomonas} that uses two
flagella, and the much studied protozoan \textit{Paramecium} which
is covered by a layer of cilia. This layer is made out of
approximately four thousands cilia which produce a very efficient
motion with a velocity of order 1$mm/s$ in water, corresponding to
$10$ times the \textit{Paramecium} size/s. Humans have ciliated
cells in several organs: in the brain (cerebrospinal fluid flow),
the retina (photoreceptor connective cilia), the respiratory tract
(epithelial cells), the ear (hair bundles), the Falopian tube or
the kidney...

Cilia have two major roles: (i) detection (sensory cilia or
flagella), for example in the retina, the ear and the kidney (ii)
propulsion or creation of fluid flow (motile cilia or flagella) as
for \textit{Paramecium} or in the respiratory tract where the
fluid flow is used to move away the mucus.

The common structure of most cilia and flagella is an axoneme
wrapped by the plasma membrane. The (9+2) axoneme is made of $9$
microtubule doublets arranged on a circle around a central pair of
microtubules \cite{Alberts}. The cilium or flagellum is attached
to the cell membrane by a basal body made of $9$ microtubule
triplets which has a structure very similar to  that of a
centriole. The basal body is attached to the cell membrane by
anchoring fibers \cite{Anderson}. Typically the radius of an
axoneme is 100 nm. The main structural difference between cilia
and flagella is their length. The typical length of a cilium is
10$\mu m$ whereas a flagellum can be ten times longer.

Dynein molecular motors are attached to the  $9$ microtubule
doublets; they move towards the microtubules $-$ ends linked to
the basal body and exert forces on the microtubules. Upon
consumption of Adenosine-Tri-Phosphate (ATP), dynein motion
generates forces that induce a sliding between adjacent
microtubules. Because the whole structure is attached at its
basis, this sliding motion induces the bending of the cilium or
flagellum and its beating.

We here focus on ciliated cells creating fluid flow. These are
cells with cilia on their surface, beating in one preferred
direction in a coordinated way.  One central feature of cilia
beating is the existence of two phases with a broken symmetry.
Each beating can be decomposed into an effective stroke (ES) that
propels the fluid and a recovery stroke (RS) where the cilium is
coming back against the flow. In the example of
\textit{Paramecium} in water, the effective stroke lasts typically
9ms whereas the recovery stroke lasts 26ms. The typical beating
frequency in water is 30$Hz$ \cite{Sleigh}. The beating of
\textit{Paramecium} cilia is 3-dimensional but for some species
like \textit{Opalina} the cilia remain in the same plane during
their beating and the beating is 2-dimensional. In this work, we
discuss the role of an external velocity field in this left-right
symmetry breaking between the effective stroke and the recovery
stroke for planar beating.

One of the most striking features of an assembly of beating cilia
is that they all beat in the same direction: the surrounding fluid
can only be propelled efficiently if all the beatings have  the
same orientation. In all mature ciliated cells, the beating
direction  is defined by the anchoring of the basal foot on the
basal body. Only newly formed or developing cilia are randomly
oriented \cite{Hagiwara}. When they start beating, they tend to
spontaneously align to finally beat in the same direction. One of
the questions addressed in this article is the nature of the
parameters that control this orientation.

The role of the central pair of microtubules in the center of the
axoneme is also a fundamental and complex question. In many
species (such as \textit{Chlamydomonas}), the central pair is both
rotating and twisting within the axoneme during the axoneme
movement. Current models postulate that the central pair modulates
dynein activity along outer microtubule doublets \cite{Porter}. It
thus allows the axoneme motion because if all the dyneins were
acting at the same time, no bending would occur. Evidence in
support of this model includes the observation that sliding
between adjacent doublets occurs preferentially along doublets
closest to one of the two microtubules of the central pair (the
C1) in \textit{Chlamydomonas} flagella  \cite{Wargo}.
Nevertheless, there exist also motile cilia with a (9+0) axoneme
having no central pair. This means that cilia beating is possible
even in the absence  of  the central pair of microtubules. Despite
its importance, we do not discuss the role of the central pair
 in the present work and we postpone its discussion to future work.

Another important feature of ciliated cells, is the existence of
waves propagating all along the surface.  These are called
metachronal waves and might be due to the coordination of adjacent
cilia for example via hydrodynamic interactions. Experimentally
metachronal waves are observed to propagate in all possible
directions: in the direction of the effective stroke (symplectic
metachronal waves), in the opposite direction (antiplectic), or
even in a perpendicular (laeoplectic or dexioplectic) or oblique
direction. The origin of these waves and the mechanisms
controlling their formation are not well understood. We show in
this article that metachronism can arise naturally from the
hydrodynamic couplings between cilia. Using a two-state model for
the dynein motion as an internal mechanism of the cilia,
metachronism appears to be a local minimum in the oscillation
threshold of the motors \cite{Camalet,Franky}.

A last important feature of cilia beating that we wish to mention,
is the role of calcium ions. The local $[Ca^{2+}]$ concentration
has a strong influence on the beating pattern of cilia or
flagella. For example detergent-treated {\it Paramecium} are able
to swim forward at low $[Ca^{2+}]$ concentration ($ <10^{-6}M$)
and backward at high $[Ca^{2+}]$ concentration ($>10^{-6}M$)
because of ciliary reversal: the directions of effective and
recovery strokes are switched \cite{Naitoh01,Naitoh02}. In any
case,  the wild type {\it Paramecium} can have a very efficient
backward motion monitored by calcium tanks in its body. We only
discuss here qualitative aspects of the role of calcium.

In this paper, we address the question of the spontaneous
alignment of an array of beating cilia and the possibility of a
spontaneous symmetry breaking in the beating that leads to the
appearance of a macroscopic fluid flow. The internal mechanism of
the cilia is described by the model of references
\cite{Franky,Camalet} which is based on a two-state model to
describe the cooperative effects between dynein motors and only
considers the relative sliding of two microtubules in the axoneme.
The coordination between the cilia is due to hydrodynamic
interactions which are discussed in details in a coarse-grained
description where the effect of the cilia on the flow is replaced
by an effective force. The outline of the paper is as follows. In
the next section, we give a simple model for the alignment of
beating cilia. In section \ref{ciliummotor}, we discuss the
beating of one cilium following the model of Jülicher and Camalet
\cite{Franky,Camalet}. Finally, in section \ref{breaking} we
discuss  the spontaneous breaking of the left-right symmetry of
the beating due to the flow created by the cilia themselves.

\section{Spontaneous alignment of an array of cilia: a simple model}
\label{rotation}
\subsection{Experimental results}

In an assembly of cilia covering the surface of a mature cell,
cilia are beating in a preferred direction, and only newly formed
or developing cilia are randomly oriented \cite{Hagiwara}. We
first discuss the experiments showing how this preferred
orientation is chosen.

As mentioned before, the ciliary axoneme grows from a basal body
analogous to a centriole. Two basal body appendages, the basal
foot and the striated rootlet, located in the axial plane of the
effective stroke, confer an asymmetrical organization to the basal
body. The basal foot is laterally associated with two consecutive
triplets and points in the direction of the effective stroke
\cite{Gibbons, Sorokin}. The striated rootlet, associated with the
proximal end of the basal body, sinks into the cytoplasm in the
opposite direction \cite{Gibbons}. These two appendages define
therefore an orientation of a cilium independent of the beating
motion.

During ciliogenesis, newly formed basal bodies migrate toward the
cell membrane where they anchor with no apparent order. Anchoring
induces axoneme assembly, and cilia grow in random orientations.
While cilia are growing, they do not beat immediately. A
reorientation by rotation of the basal bodies in a common
direction occurs at the final stage of ciliogenesis, when mature
cilia beat \cite{Boisvieux}. The preferred direction of the
assembly is then well-defined. In the immotile-cilia syndrome,
axonemes are incomplete, and the ciliary activity is abnormal or
absent: the fluid is poorly or not propelled. On the cell level,
the basal bodies are randomly oriented \cite{Afzelius}.

These experimental facts suggest that the beating and orientation
of cilia are closely related. Our working hypothesis is that the
alignment of an assembly of beating cilia is mostly due to
hydrodynamical coupling between cilia. The global flow created by
the other cilia tend to orient a given cilium and above a certain
beating amplitude, all cilia orient in the same direction. We now
give a very simple modelling of this cooperative alignment.

\subsection{Alignment transition}

We assume in the following that the beating is planar. It is the
case for \textit{Opalina} for example, but not exactly for
\textit{Paramecium} where the recovery stroke is not in the plane
of the effective stroke.

\begin{figure}
\begin{center}
\includegraphics[angle=0, width=9cm]{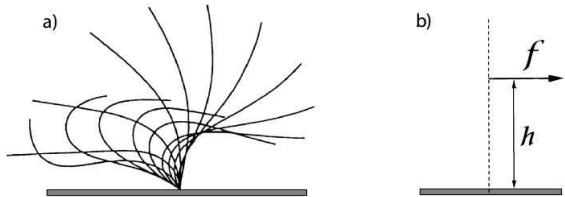}
\end{center}
\caption{\label{Beating}a) Beating pattern of a single cilium
showing the Effective Stroke (ES), where the fluid is efficiently
propelled, and the slower Recovery Stroke (RS),  where the cilium
comes back close to the surface to minimize the viscous effects.
b) Effective force in the fluid $ \vec{f} $ applied at a height $
h $ above the cell membrane, to mimic the cilium beating.}
\end{figure}

Near the top of the ciliary layer, observations show that the
velocity is time independent and uniform \cite{Liron}.
Consequently, we average the beating over one time period and
replace each cilium of length $L$ (and its effective and recovery
stroke) by a single force (stokeslet) $ \vec{f} $, parallel to the
surface, created in the fluid of viscosity $ \eta $ at height $
h<L $ above the membrane, as sketched in figure \ref{Beating}.

We choose the $ x $ axis in the direction of the effective stroke
and the $z$ axis perpendicular to the cell surface that we
approximate by a plane. The stokeslet along the x axis
($\vec{f}=f\vec{e_{x}} $), is located at point $ \vec{S}=(0,0,h)$.

The velocity created at point $ \vec{X}=(x,y,z) $ by this stokeslet with
a no-slip boundary condition on the plane $ z=0 $ is given by :
\begin{equation}
\nonumber \vec{v}(\vec{X})=\textbf{G}(\vec{X},\vec{S}).\vec{f}
\end{equation}
where the response tensor tensor $ \textbf{G}$ is given in \cite{
Blake} and reads:
\begin{eqnarray}
\nonumber 8\pi\eta \ \textbf{G}_{jk}(\vec{X},\vec{S})=\left(
\dfrac{\delta_{jk}}{\rho}+\dfrac{\rho_{j}\rho_{k}}{\rho^{3}} \right) -\left(
\dfrac{\delta_{jk}}{R}+\dfrac{R_{j}R_{k}}{R^{3}} \right)
\\
\label{Gtensor}
 +2S_{z}\left(\delta_{k\alpha}\delta_{\alpha
l}-\delta_{kz}\delta_{zl}\right) \dfrac{\partial}{\partial
R_{l}}\left(\dfrac{S_{z}R_{j}}{R^{3}}
-\dfrac{\delta_{jz}}{R}-\dfrac{R_{j}R_{z}}{R^{3}}\right)
\end{eqnarray}
where $ \vec{\rho}=\vec{X}-\vec{S}$, $\vec{R}=\vec{X}+\vec{S}$ and
$\alpha=x,y$.

In order to simplify the hydrodynamic problem, we assume in all
the following that the cilia are far away from each other. In this
asymptotic limit, it is consistent to describe the effect of the
cilium in the fluid by a stokeslet. We introduce the
two-dimensional vector along the cell surface $ \vec{r}=(x,y) $
and consider the limit  $ r \gg z,h,L $. We are interested in the
velocity in the vicinity of the ciliary layer, typically $\
0<z<1.5 \ L $. At lowest order in $ z/r $, the velocity reads:
\begin{equation}
\label{vitessev} \vec{v}(r,\theta ,z)=\frac{3fh}{2\pi \eta}
\frac{z\cos\theta}{r^{3}} \vec{e_{r}}+ \mathcal{O}(z^{3}/r^{3})
\simeq \vec{u}(r,\theta)\ \bar{z}
\end{equation}
where we have defined a dimensionless height $\bar{z}=z/L $. In
the following, we use mostly the velocity $ \vec{u}(r,\theta) $.
At this order, the flow field is a radial flow centered on the
cilium-stokeslet. Note that because of the no slip boundary
condition on the surface, the force appears in the velocity field
(Eq. \ref{vitessev}) in the combination $ fh $ homogeneous to a
momentum.

\begin{figure}
\begin{center}
\includegraphics[angle=0, width=7cm]{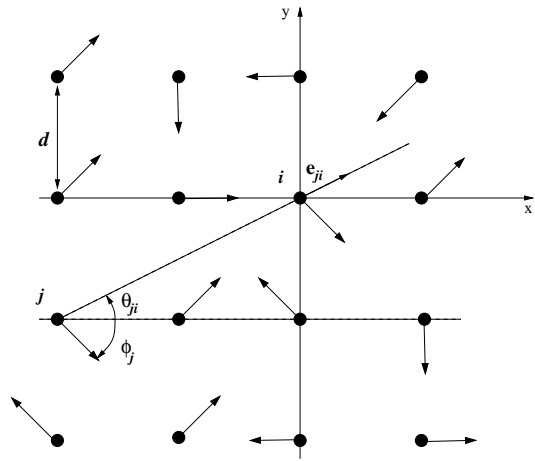}
\end{center}
\caption{\label{Lattice} Square lattice of cilia with a distance $
d $ between two neighboring cilia. Cilium $ j $ exerts on the
fluid a force $ f $ in the direction $ \phi_{j} $; $ \theta_{ji}=(
\vec{e_{x}},\vec{e_{ji}}) $ where $ \vec{e_{ji}} $ is the unit
vector from cilium $ j $ to cilium $ i $.}
\end{figure}

We consider now a regular array of cilia; this is a reasonable
assumption for \textit{Paramecium}, which shows a beautiful and
very regular array of cilia on its surface \cite{Beisson}. We
assume that this array is in an infinite plane. Cilium $ i $ is
defined by its position in the $x,y$ plane (the cell surface) by a
vector $ \vec{r_{i}}$ and by the angle of its plane of beating
with the $x$ axis, $ \phi_{i}$ as displayed on Fig. \ref{Lattice}.

The total velocity at the cilium $ i $ at height $z$, $
\vec{V}(\vec{r}_{i},z) $ is the sum of all the velocities $
\vec{v}_{j}(\vec{r}_{i},z) $ created by the other cilia $ j \neq
i$:
\begin{equation}
\nonumber \vec{V}(\vec{r}_{i},z)=\sum_{j\neq i}
\vec{v}_{j}(\vec{r}_{i},z)
\end{equation}
We single out the the $\bar{z}$ dependence of $\vec{V}$ writing $
\vec{V}(\vec{r}_{i},z)= \vec{U}(\vec{r}_{i})\ \bar{z}$ with
$\vec{U}(\vec{r}_{i})=\sum_{j\neq i} \vec{u}_{j}(\vec{r}_{i})$ and
\begin{equation}
\label{vitesseqcq} \vec{u_{j}}(\vec{r_{i}})\simeq\frac{3fhL}{2\pi
\eta} \frac{\cos(\theta_{ji}-\phi_{j})}{\vert
\vec{r_{i}}-\vec{r_{j}} \vert^{3}} \vec{e_{ji}}
\end{equation}
where $ \vec{e_{ji}} $ is a unit vector from cilium $ j $ to
cilium $ i $ and $ \theta_{ji}=( \vec{e_{x}},\vec{e_{ji}}) $, as
shown on Fig. \ref{Lattice}.

We now use a mean field approximation, replacing the velocity $
\vec{u}_{j}(\vec{r}_{i})  $ by its average over the directions $
<\vec{u}_{j}(\vec{r}_{i})>_{\phi} $ given by:
\begin{equation}
\label{vitessemoy} <\vec{u_{j}}(\vec{r}_{i})>_{\phi}=
\int_{0}^{2\pi} d \phi \ P(\phi)\ \vec{u_{j}}(\vec{r}_{i})
\end{equation}
where $ P(\phi) $ is the probability for a cilium to make an angle
$ \phi $ with the $ x $ axis. The average velocity at cilium $i$ is
$ \vec{U}(\vec{r}_{i})\simeq \sum_{j\neq i} <\vec{u}_{j}(\vec{r}_{i})>_{\phi}$.

The mean field approximation assumes that the fluctuations around
a given angle determining the direction of the flow are small. We
choose the $ x $ axis in the direction of the flow without any
loss of generality, so that the probability $ P(\phi) $ is peaked
around $\phi=0 $.

In order to determine the probability $ P(\phi) $, we write a
stationary Fokker-Planck equation $\partial_{\phi} J=0$. The
probability current $J=P \partial_{t} \phi - D_{r}\partial_{\phi}
P$ is the sum of two terms: a convection term and a diffusion term
where $ D_{r} $ is a rotational diffusion coefficient. The beating
plane can fluctuate due to thermal fluctuations. Because of the
flow, if the beating plane of one cilium is at an angle $ \phi $
with the flow direction, the cilium is subject to a torque $
M_{z}^{flow}= -\alpha U \sin \phi  $ along the $ z $ axis that
tends to align it in the direction of the flow, where $ U \bar{z}
$ is the velocity of the global flow and $\alpha$ a viscous
coefficient involving the geometry of the cilium. A rotating
cilium is also subject to a viscous torque $ M^{viscous}$ opposing
the rotation $M_{z}^{viscous}= -\zeta \partial_{t} \phi $ where $
\zeta $ is the rotational friction constant. The total torque on
the cilium vanishes ($M_{z}^{flow}+M_{z}^{viscous}=0$) and the
probability distribution satisfies the Fokker-Planck equation
\begin{equation}
\nonumber \dfrac{\partial^{2} P}{\partial \phi ^{2}}+
\dfrac{\alpha U }{D_{r}\zeta} \dfrac{\partial}{\partial \phi}[P
\sin \phi ]=0
\end{equation}
Defining the effective temperature as  $ D_{r}\zeta = k_{B}T $ and imposing
the normalization condition $ \int P(\phi)d \phi = 1 $, we obtain:
\begin{equation}
\label{proba} P(\phi)=\dfrac{e^{\frac{\alpha U}{k_{B}T}\cos
\phi}}{2 \pi I_{0}(\frac{\alpha U }{k_{B} T})}
\end{equation}
where  $I_{0}(x)$ is the modified Bessel function defined in
\cite{grad}. The average velocity $\vec U $  can then be
self-consistently determined by calculating $
<\vec{u}_{j}(\vec{r}_{i})>_{\phi} $, and summing over all the
lattice sites. We obtain
\begin{equation}
\label{self} \vec{U}= \dfrac{3fhL}{2 \pi \eta d^{3}}\   K \
\dfrac{I_{1}(\frac{\alpha U }{k_{B}T})}{I_{0}(\frac{\alpha U
}{k_{B}T})} \ \vec{e_{x}}
\end{equation}
where $ K $ is a constant depending on the nature of the lattice,
$d$ is the lattice constant (the distance between cilia) and $
I_{1}$ is a modified Bessel function \cite{grad}. For a square
lattice,  \cite{internet}:
\begin{equation}
\label{constanteK} K_{square}=\sum_{(k,l)\neq (0,0)}
\dfrac{k^{2}}{(k^{2}+l^{2})^{5/2}}=2 \  \beta (\frac{3}{2}) \zeta
(\dfrac{3}{2})\simeq 4.52
\end{equation}
$ \beta (s) $ and  $ \zeta (s) $ being respectively the Dirichlet
and the Riemann functions \cite{grad}.

The self-consistent equation for the flow velocity can be
discussed by expanding the integrals $ I_{0} $ and $ I_{1} $ in
the vicinity of $ U=0 $: there are two solutions $U=0$, and a
solution at a finite velocity which exists only within a certain
range of parameters
\begin{equation}
\label{flow} U=2\sqrt{2} \dfrac{k_{B}T}{\alpha} \
\sqrt{1-\dfrac{4\pi k_{B}T  \eta d^{3}}{3K\alpha fhL}}
\end{equation}
This solution exists only if
\begin{equation}
\label{condition}
\dfrac{3K\alpha fhL}{4\pi k_{B}T \eta d^{3}}\ >1
\end{equation}
Within the mean field approximation Eq. \ref{condition} defines a
dynamical phase transition between a non-moving fluid with
randomly oriented cilia and a moving fluid with a global flow $
V(z)=U\bar{z}\neq 0 $ given by Eq. \ref{flow} where all cilia are
spontaneously aligned in the same orientation. This dynamical
phase transition is second order (with a continuous velocity at
the transition) and it is associated to a spontaneous breaking of
the initial $ O(2) $ symmetry.

The influence of some of the parameters can be directly analyzed
on Eq. \ref{condition}. A decrease of the distance $ d $ between
two cilia favors the alignment, increasing the hydrodynamic
coupling. A decrease of the temperature $T$ also favors alignment
as the random thermal motion opposes the alignment. An increase of
the effective hydrodynamic force of one cilium  $f$ is associated
to an increase the hydrodynamic interactions between cilia and
leads to a better alignment. The same effect occurs for $ \alpha $
and the cilium length $L$. Finally, increasing $ h $ helps to
create a global flow, since the velocity on the membrane vanishes
and the higher the force is exerted, the more efficient.

A more precise analysis requires the estimation of the parameters
$ f, \alpha $ and $ h $. The height is of the order of the cilium
size $h \sim L$. The calculation of $ \alpha $ is given in
appendix I for a general beating (Eq. \ref{alpha}). It turns out
that $\alpha $ is linked to the difference of the areas covered
during the effective and recovery strokes. Here we approximate $
\alpha \sim \xi_{\perp} L \mathcal A $, where $\xi_{\perp}$ is the
perpendicular friction constant per unit length of the cilium and
$ \mathcal A$ the amplitude of the movement of the tip.

In order to give a simple estimation of  the effective
hydrodynamic force, we consider that the friction coefficient
takes the perpendicular value $ \xi_{\perp}$ during the effective
stroke, and the parallel value $\xi_{\parallel}$ during the
recovery stroke. Introducing the beating frequency  $ \omega $, we
estimate $f \sim(\xi_{\perp}-\xi_{\parallel})L \omega \mathcal A
$. A more precise calculation of these two quantities as a
function of the beating patterns is given in appendix I.

Consequently, we obtain:
\begin{equation}
\nonumber \dfrac{3K\alpha fhL}{4\pi k_{B}T \eta d^{3}}\ \sim
\dfrac{\xi_{\perp}(\xi_{\perp}-\xi_{\parallel})\mathcal A^{2}L
\omega}{k_{B}T \eta}\dfrac{L^{3}}{d^{3}}
\end{equation}
 The difference
between the two local drag coefficients $ \xi_{\perp} $ and $
\xi_{\parallel} $ is the key to an efficient beating. Increasing
the amplitude $\mathcal A$ or the frequency $ \omega $ of the
beating favors the alignment of cilia, as could be expected.
Increasing the viscosity of the medium also promotes the
transition, because it increases the coupling between cilia.

This naive mean field approximation is only a first step of our
study. In the following, we take a closer look at the  internal
beating mechanism of one cilium and then at the beating of an
array of cilia to obtain a more precise and quantitative
description.

\section{Axonemal beating}
\label{ciliummotor}
 In this section we discuss the beating mechanism of a single cilium.
We follow closely the work of Camalet and Jülicher  \cite{Franky}
which mimics the cilium by two microtubule filaments sliding along
one another under the action of the dynein motors and  uses a
2-state model to describe the collective motion of the dyneins. We
use as boundary conditions for the motion those introduced
recently by Hilfinger and Jülicher \cite{Andy} that seem to have
good experimental support \cite{Geraint}. In the next section, we
use the same model to discuss the coordination between cilia.

\subsection{Equation of motion}

\begin{figure}
\begin{center}
\includegraphics[angle=0, width=6cm]{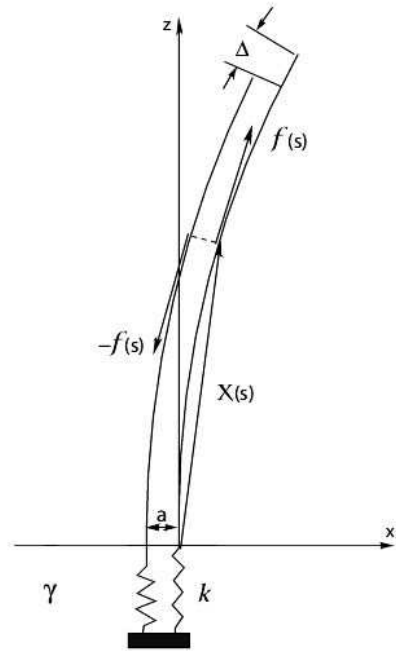}
\end{center}
\caption{\label{Filaments}Two filaments (full curves) $ \vec{X} $
and $ \vec{X}' $ at constant separation $ a $ are rigidly
connected at the bottom end where $ s=0 $. Internal forces $ f(s)
$ are exerted in opposite directions, tangential to the filaments.
The displacement $ \Delta $ at the tip is indicated.}
\end{figure}

Each microtubule doublet within the axoneme can be described
effectively as an elastic rod. Deformations of this rod lead to
local sliding displacements of neighboring microtubules. Here, we
only consider planar deformations. In this case the geometrical
coupling between bending and sliding can be captured by
considering two parallel elastic filaments (corresponding to two
microtubule doublets) with a constant separation $a$ along the
whole length of the rod (see Fig. \ref{Filaments}). At one end,
which corresponds to the basal end of an axoneme, the two
filaments are elastically attached and are allowed to slide with
respect to each other, but not to tilt \cite{Andy}. The basal
connection is characterized by an elasticity $ k $ and a
frictional drag $ \gamma $. The configurations of the axoneme are
described by the shape of the filament pair given by the position
of one filament $ \vec{X}(s) $ at arclength $s$. The shape of the
other filament is then given by $ \vec{X'}(s)= \vec{X}(s)
-a\vec{n}(s) $, where $ \vec{n} $ is the filament normal. In the
following, we describe the filament conformation by the angle
$\psi$ between the local tangent vector and the $z$ axis or by the
deformation $h$ in the transverse direction defined in Fig.
\ref{Notations}.

The energetics of the filament pair is due to the bending
elasticity. In addition to filament bending, we also take into
account internal stresses due to the active elements (dyneins). We
characterize them by the force per unit length $ f(s) $ acting at
position $ s $ in opposite directions on the two microtubules.
This force density corresponds to a shear stress within the cilium
which tends to slide the two filaments with respect to each other.

The local curvature is $ C =\partial_{s}\psi$ (see Fig.
\ref{Notations}). The sliding displacement, $ \Delta(s,t) $ is
related to the  the sliding displacement at the base $ \Delta_{0}
(t) $ by $\Delta (s,t)-\Delta_{0} (t) = a \psi(s,t)$, because we
impose the boundary condition $ \psi(0,t)=0$.

A configuration of a filament pair of length $ L $ is associated
to the free energy functional:
\begin{equation}
\nonumber G=\dfrac{k}{2}\Delta_{0}^{2} + \int_{0}^{L}
ds[\frac{\kappa}{2}C^{2} - f \Delta + \frac{\Lambda}{2}
(\partial_{s}\vec{X})^{2}]
\end{equation}
Here, $\kappa$ denotes the total bending rigidity of the
filaments. The inextensibility of the filaments is taken into
account by the Lagrange multiplier $ \Lambda(s) $ which enforces
the constraint $ (\partial_{s}\vec{X})^{2}=1 $.

The first term of this equation is the elastic energy due to the
basal sliding occurring with a connection of elasticity $ k $.

The tangent component of the integrated forces acting on the
filament between $ s $ and $ L $ is denoted by $ \tau(s) $.
Assuming that there is no external force applied at the end $s= L$
of the cilium:
\begin{equation}
\nonumber \tau(s)=\vec{t}(s).\int_{s}^{L} ds' \ \frac{\delta
G}{\delta \vec{X}}=-\vec{t}(s). \int_{0}^{s} ds' \ \frac{\delta
G}{\delta \vec{X}}
\end{equation}

We assume for simplicity, that the hydrodynamic effects of the
surrounding fluid can be described by two local friction
coefficients $ \xi_{\perp} $ and $ \xi_{\parallel} $ for normal
and tangential motion. The total friction force per unit length
exerted by the cilium on the fluid is then
$\vec{f_v}[\vec{X}(s)]=(\xi_{\parallel}\vec{t}\
\vec{t}+\xi_{\perp}\vec{n}\ \vec{n})\ \partial_{t}\vec{X}(s)$. The
force balance at arclength $s$ can then be written as
\begin{equation}
\label{Eq}
\partial_{t}\vec{X}=-(\frac{1}{\xi_{\perp}}\vec{n}\
\vec{n}+\frac{1}{\xi_{\parallel}}\vec{t}\ \vec{t})\ \frac{\delta
G}{\delta \vec{X}}
\end{equation}
which leads to:
\begin{equation}
\label{flagelle}
\partial_{t}\vec{X}= \frac{\vec{n}}{\xi_{\perp}}(-\kappa \dddot {\psi} - a \dot f + \dot \psi \tau)
+ \frac{\vec{t}}{\xi_{\parallel}}(\kappa \dot \psi \ddot \psi + a \dot \psi f + \dot \tau)
\end{equation}
where the derivatives with respect to arclength have been denoted
by a dot.

The beating of the filaments is very sensitive to  the boundary
conditions imposed at its ends. As the the force density $-\delta
G/\delta \vec{X} $ is equilibrated by the density of friction
force exerted by the fluid on the system (see Eq. \ref{Eq}), the
boundary contributions coming from the free energy variation $
\delta G $ are equilibrated by external forces $ \vec{f}_{ext} $
and torques $ \vec{T}_{ext}= T_{ext}\ \vec{e}_{y} $ applied at the
ends.

At the free end of the cilium, both the external force and the
external torque vanish and
\begin{equation}
\label{limit1}
\vec{f}_{ext} = -\left( \kappa \dot C + a f
\right)\vec{n} +  T \vec{t} =\vec 0
,\quad
T_{ext} = \kappa C =0
\end{equation}
At the base, $s=0$, the boundary conditions  are:
\begin{equation}
\label{limit2} \vec{f}_{ext} = \left( \kappa \dot C + a f
\right)\vec{n} -  T \vec{t}
,\quad  T_{ext} = -\kappa C + a \int_{0}^{L}d s f(s)
\end{equation}
The external torque and force are chosen in such a way that the
base is fixed ($ \partial_{t}\vec{X}=\vec{0} $) and that the
cilium remains perpendicular to the surface (\
$\partial_{t}\vec{t}=\vec{0} $ or $\psi(0)=0$).

The final boundary condition is associated to the basal sliding
\begin{equation}
\label{basalsliding} \gamma \partial_{t} \Delta_{0}=-\dfrac{\delta
G}{\delta \Delta_{0}}=-k \Delta_{0} + \int_{0}^{L} ds f(s)
\end{equation}

The determination of the cilium motion requires a model for the
shear force created by the dyneins that we calculate using a
two-state model.

\subsection{2-state model for the cilium}

Following \cite{Franky}, we now introduce the 2-states model of
coupled molecular motors \cite{Prost, Prost1} to describe the
internal mechanism of the cilium.  This model allows the
calculation of the shear force $f$ due to the dyneins.

Each motor has two different chemical states, a strongly bound
state, 1, and a weakly bound state, 2. The interactions between a
motor and a filament in both states are characterized by potential
energy landscapes $ W_{1}(x) $ and $ W_{2}(x) $, where $ x $
denotes the position of a motor along the filament. The potentials
have the filament symmetry: they are periodic with period $ l $, $
W_{i}(x)= W_{i}(x+l) $ and are, in general, spatially asymmetric,
$ W_{i}(x)\neq W_{i}(-x) $.

In the presence of ATP, the motors undergo transitions between
states with transition rates $ \omega_{1} $ and $ \omega_{2} $.
Introducing the relative position $ \xi $ of a motor with respect
to the potential period, ($ x=\xi+nl $ with $ 0\leq\xi<l $ and $ n
$ an integer), we define the probability $ P_{i}(\xi,t) $ for a
motor to be in state $ i $ at position $ \xi $ at time $ t$. The
relevant Fokker-Planck equations are:
\begin{eqnarray}
\nonumber \partial_{t}P_{1}+ v \
\partial_{\xi}P_{1}&=&-\omega_{1}P_{1}+ \omega_{2}P_{2}
\\
\nonumber \partial_{t}P_{2}+ v \
\partial_{\xi}P_{2}&=&\omega_{1}P_{1}- \omega_{2}P_{2}
\end{eqnarray}
where $ v=\partial_{t}\Delta=a \partial_{t}\psi(s)$ is the sliding
velocity between the 2 filaments.

The simplest choice of the two potentials $ W_{i} $, is a
saw-tooth potential (with barrier height $ U \gg kT $)
representing a strongly bound state for $ W_{1} $, and a flat
potential $ W_{2} $ representing a weakly bound state. Here, for
simplicity, we use the symmetric potentials:
\begin{eqnarray}
\nonumber W_{1}(x)&=&U \sin^{2}(\pi \frac{x}{l}) \\
\nonumber W_{2}(x)&=& W_{2}
\end{eqnarray}
Although this choice is somehow arbitrary, we checked that the
final results only depend qualitatively on the actual shape of the
potentials and of the rates defined below.

When a number of motors act together to propel a filament,
however, the direction of motion is a collective property. The
filament might move in either direction \cite{Prost1}. The absence
of asymmetry in the potentials implies that an individual motor is
not able to move directionally. It is not the case for an assembly
of motors: even with a symmetric potential, provided that
detachment can only take place at a localized position near the
bottom of a potential well, oscillations can occur.

We define the distance from equilibrium $ \Omega $:
\begin{equation}
\Omega=Sup_{[0,l[}\mid
\frac{\omega_{1}}{\omega_{2}}-e^{\frac{W_{1}-W_{2}}{kT}}\mid \
\propto e^{\Delta\mu / kT}-1
\end{equation}
$ \Omega $ is related to the chemical potential difference between
ATP and its hydrolysis products, $ \Delta\mu  =
\mu_{ATP}-\mu_{ADP}-\mu_{P} $. At equilibrium, $ \Delta\mu  =0 $
and $ \Omega=0 $. We assume for simplicity that the binding rates
$ \omega_{2} $ and the detachment rate $ \omega_{1} $ are given
by:
\begin{eqnarray}
\nonumber \omega_{2}(\xi)&=&\nu (1+\Omega \sin ^{2} (\pi\frac{\xi}{l}))\\
\nonumber \omega_{1}(\xi)&=&\nu\Omega \cos ^{2}(\pi \frac{\xi}{l})
\end{eqnarray}
Note that, with this choice the sum $\nonumber
\omega_{1}+\omega_{2}=\nu(1+\Omega)$ does not depend on $ \xi $,
and that if  $\Omega=0 $, $\omega_{1}= 0 $ and no directional
movement is possible. Here $\nu $ is a constant transition rate.

If we assume that the motors are uniformly distributed along the
filaments with a density $ \rho $, the probabilities $ P_{1} $ and
$ P_{2} $ satisfy the relationship $ P_{1}+P_{2}=\rho$. The Fokker
Planck equation reduces then to a single equation for $P=P_1$
\begin{equation}
\label{Proba}
\partial_{t}P+ (\partial_{t}\Delta) \partial_{\xi}P=-(\omega_{1}+\omega_{2})P+\rho \omega_{2}(\xi)
\end{equation}

This model leads to an expression for the shear force per unit of
length $ f(s,t) $ created by the dyneins and driving the the
cilium beating. Using the results of \cite{Prost}, and the fact
that $ W_{2} $ is a constant:
\begin{equation}
\label{shearforce} f(s,t)=-K\Delta -\lambda \partial_{t}\Delta
-\frac{1}{l}\int_{0}^{l}d\xi\ P(\xi)\partial_{\xi}W_{1}
\end{equation}
where $ K $ is an elastic stiffness per unit  length mimicking the
influence of the nexins which are proteins  acting as springs in
the axonemal structure, and $ \lambda $ is an internal friction
coefficient per unit length modelling the friction encountered by
the motors. Equations \ref{flagelle} and \ref{Proba} allow in
principle a complete calculation of the beating motion.

In the following, we assume that the beating occurs with a "small"
amplitude which means that both $\psi (s,t) \ll  1$ and $ h(s,t)
\ll  L$. A quick look at the beating pattern of a cilium of
\textit{Paramecium} shows that the beating occurs with a large
amplitude. Nevertheless, this approximation allows us to extract
interesting information on the parameters controlling the beating.
Moreover, the work of Hilfinger shows that larger amplitude
beating patterns are very similar to small amplitude patterns \cite{Andy}. We
must however keep in mind that our approach is valid only if the
the system stays close to an oscillation bifurcation which is
consistent with the fact that we consider only small movements.

We use the deformation $ h $, rather than $ \psi $ or $ \Delta $
to describe cilium motion and work at second order in $ \vert
h \vert $ so that $\psi = \dot h + \mathcal{O} (\vert h \vert^{3}
)$. In the absence of any external flow, the equation of motion
\ref{flagelle} projected on $ \vec{t} $ imposes that $\tau =
\mathcal{O} (\vert h \vert^{2})$. The projection of the equation
of motion on $ \vec{n} $ then yields:
\begin{equation}
\label{flagelle1} \xi_{\perp} \partial_{t}h = -\kappa \ddddot h -
a \dot f + \mathcal{O}(\vert h \vert^{3}  )
\end{equation}
The non-linear terms are not important here but they will turn out
important in the following section. Indeed, experiments show that
the beating is clearly asymmetrical (ES and RS), and we must
expand at least to next order if we want to capture this
phenomenon. With this variable and at this order, the boundary
conditions read:
\begin{eqnarray}
\label{bound}
h(s=0)=0, \quad \dot {h}(s=0)=0 \nonumber  \\
\kappa \dddot{h}(s=L)+a f(s=L)=0, \quad   \ddot {h}(s=L)=0
\end{eqnarray}

The explicit solution of the equation of motion \ref{flagelle1} is
obtained by Fourier expansion in time
$h(s,t)=\sum_{n=-\infty}^{\infty}  h_{n}(s) e^{i n\omega t}$. The
explicit derivation of the equation satisfied by the Fourier
components is given in appendix II. At linear order the effect of
the motors is characterized by a susceptibility
\begin{equation}
\label{chi} \chi(\Omega,\omega)=-K-\lambda i \omega + \dfrac{\pi^{2} \rho U}{2 l^{2}}
\dfrac{i \Omega  \omega}{(1+\Omega)((1+\Omega)\nu+i\omega)}
\end{equation}
Using dimensionless variables,  $\bar s=s/L$, $\bar
\omega=\dfrac{\xi_{\perp}L^{4} }{\kappa}\omega$,
 $\bar \chi _{n} =
\dfrac{a^{2}L^{2}}{\kappa}\chi(\Omega, n \omega)$ and $\bar h=h/L$,
 the equation of motion for the Fourier component $n$ reads
\begin{equation}
\label{flagelle3}
\ddddot{\bar{h}}_{n}+\bar\chi_{n}\
\ddot{\bar{h}}_{n}+in\bar \omega\ \bar{h}_{n}=0
\end{equation}
The boundary conditions at the base $ \bar s=0 $ are
\begin{equation}
\label{bound0}
\bar{h}_{n}(0)=0 , \quad \dot{\bar{h}}_{n}(0)=0
\end{equation}
At the free end of the cilium, $ \bar s=1 $ :
\begin{eqnarray}
\label{bound1}
\dddot{\bar{h}}_{n}(1)+\bar\chi_{n}
\dot{\bar{h}}_{n}(1)+ \bar \Gamma_{n}{{\bar{h}}_{n}}(1)=0
,\quad \ddot{\bar{h}}_{n}(1)=0
\end{eqnarray}
with
$ \bar \Gamma_{n}={\bar\chi_{n}^{2}}
/\left( \bar k-\bar\chi_{n}+ in\bar\gamma \bar \omega \right)$
where we have introduced the dimensionless parameters
$ \bar k=\dfrac{a^{2}L}{\kappa}k$
and $\bar \gamma = \dfrac{a^{2}}{L^{3}\xi_{\perp}}\gamma$.

\subsection{Beating pattern}

In the absence of any external flow the beating is symmetric and
the Fourier components n=0 and n=2 of $h$ vanish for symmetry
reasons. Close to the oscillation bifurcation threshold, cilium
beating is dominated by the first Fourier component $n=1$.

The solution of the linear equation of motion \ref{flagelle3} is
written as the sum of $4$ exponentials
\begin{equation}
\label{h}
{\bar{h}}_{1}(\bar s)={\cal A}_{1}\left( e^{q_{1}\bar
s}+b_{1}e^{-q_{1}\bar s}+c_{1}e^{q_{2}\bar s}+ d_{1}e^{-q_{2}\bar
s} \right)
\end{equation}
where the two inverse decay lengths are given by
\begin{eqnarray}
\label{sol1} \nonumber
 q_{1}=\left( -\dfrac{\bar\chi_{1}}{2}+
\dfrac{1}{2}(\bar\chi_{1}^{2}-4i\bar \omega)^{1/2} \right)^{1/2}
\\
 q_{2}=\left( -\dfrac{\bar\chi_{1}}{2}-
\dfrac{1}{2}(\bar\chi_{1}^{2}-4i\bar \omega)^{1/2} \right)^{1/2}
\end{eqnarray}

The boundary conditions are explicitly discussed in appendix II.
The condition for existence of non vanishing solutions is given by
equation \ref{det} of Appendix II. This is a complex equation,
that gives therefore two conditions which determine both the the
critical value of the distance from equilibrium where the
oscillations start $ \Omega_{c}$, and the reduced oscillation
frequency $\bar \omega_{c} $. The critical value $ \Omega_{c}$ is
a Hopf bifurcation threshold: there are no oscillations if $
\Omega  \leq \Omega_{c}$ and cilium
beating is only possible if $ \Omega  \geq \Omega_{c}$. At the
bifurcation threshold, the amplitude of the oscillations vanishes.
It is not possible to calculate the amplitude of the oscillations
above the bifurcation threshold with the linear theory presented
here. This requires a complete determination of the third order
terms in the equation of motion which goes far beyond the scope of
this work. This very complex problem is attacked in the work of
Hilfinger and Jülicher \cite{Andy}.

An analytical determination of the Hopf bifurcation threshold and
of the beating frequency at the threshold does not seem to be
possible analytically. We therefore rely on a numerical solution,
choosing reasonable values of the various parameters.

We study  a cilium of length $L=12\mu m$, which is the length of a
\emph{Paramecium} cilium. It moves in a fluid of viscosity $\eta
\sim 4 \eta_{water}=4\  10^{-3}Pa.s$, which is higher than the
water viscosity in order to take into account the proteins  above
the cell body.  We estimate $\kappa= 4\ 10^{-22}J.m$ corresponding
to 20 microtubules \cite{Ishijima}. The other parameters are $
l=10 nm$, $a=20nm$, $ U=10kT $, $ K \simeq 0 $, $ \rho= 5\ 10^{8}
m^{-1}$, and $ \lambda =1 Pa.s $ very similar to those of
\cite{Camalet}. In order to match the typical frequency observed
in \emph{Paramecium}, and to obtain a realistic pattern of the
beating, we take $ \xi_{\perp}=35.5 \eta=142\ 10^{-3} Pa.s$ and we
choose $ k=6.54 N m^{2}$, $ \gamma=7.31\ \eta_{water} $, and $
\nu= 600 s^{-1}$.

The value of $\xi_{\perp}$ is rather high, but it must include the
hydrodynamic interactions of the cilium with the cell surface (see
\cite{Murase}) which are not taken into account if we use the
classical friction per unit length of a rod.

\begin{figure}
\begin{center}
\includegraphics[angle=90, width=3cm]{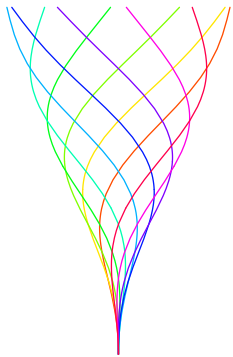}
\end{center}
\caption{\label{Calculated Beating} Approximate cilium deformation
$ \bar{h}(\bar s,t) $ at different times steps (corresponding to
different colors) during a beating period. The beating is
symmetrical with respect to the vertical axis. Deformations are
propagating from base to tip. With $ \mathcal{A}_{1}=1/70 $ the
maximum deformation is $ \bar{h}_{max}\simeq 0.14 $.}
\end{figure}

The numerical resolution of Eq. \ref{det} then yields $
(\Omega_{c},\bar \omega_{c})\simeq (6.55\ 10^{-9},1294)$. This
corresponds to a critical beating frequency $ f_{c}\simeq 28 Hz $
which is the typical value for \emph{Paramecium} \cite{Sleigh}.
The calculated beating pattern of the cilium is shown on figure
\ref{Calculated Beating}. It corresponds to a wave (or a
superposition of waves) propagating from the base of the cilium to
the free tip as observed experimentally.

A detailed study shows that the equation giving the bifurcation
threshold and the beating frequency has several solutions.
$(\Omega_{c}^{(n)},\bar \omega_{c}^{(n)}) $ with $
\Omega_{c}^{(n+1)}> \Omega_{c}^{(n)}$. A first guess would be that
the axoneme starts beating at the lowest threshold
$(\Omega_{c}^{(1)},\bar \omega_{c}^{(1)}) $. However, it is known
experimentally that during the beating the deformation waves
propagate from the base to the tip and not from the tip to the
base \cite{cells.de}. The first two oscillating modes correspond
to waves propagating in the opposite direction. In order to be
consistent with the experiments we do not consider them here. A
better choice of the transition rates $ \omega_{1} $ and $
\omega_{2} $, would perhaps allow to justify this choice. The
direction of propagation of the wave is extremely sensitive to the
boundary conditions. We have allowed here basal sliding as
suggested by some experiments and we have imposed that the cilium
is clamped at its base with an angle $\psi=0$. This also seems
consistent with some experiments analyzed in \cite{Andy}. The
other extreme limit of a completely free cilium (a vanishing
external torque at the base) leads to a wave propagating form the
base to the tip for the first mode. We have tried to use an
intermediate boundary condition where the torque at the base is an
elastic torque and varying the related stiffness, however we were
not able to obtain a beating pattern looking like the experimental
one. We therefore proceed, considering only the third beating
mode.

The beating frequency $ f_{c} $ varies with the viscosity of the
medium. Experimentally, when methyl-cellulose is added in water,
the viscosity increases significantly. We predict here a decrease
of $ f_{c} $ with increasing external viscosity  as observed in
the experiments of \cite{Machemer} and in numerical simulations
\cite{Gueron01} (see Table \ref{tableau}). We observe an
approximate linear decrease of the beating frequency when plotted
against $ log (\eta/\eta_{w})$ as in the simulations performed in
\cite{Gueron01} (we find similar values for the frequency).

\begin{table}
\begin{center}
\begin{tabular}{c c c}
External viscosity\ \ \  & \ \ \ Critical frequency $ f_{c} $\ \ \  &\ \ \  Simulations\\
\hline
$ \eta_{w} $ & $ 28 Hz $ & $ 29 Hz $ \\
\hline
$ 2\eta_{w} $ & $19 Hz $ & $ 17 Hz $\\
\hline
$ 3\eta_{w} $ & $14 Hz $ & $ 12 Hz $\\
\end{tabular}
\caption{\label{tableau} Decrease of the beating frequency with
increasing external viscosity as observed in experiments
\cite{Machemer}. Comparison with the simulations done in
\cite{Gueron01} for one single cilium.}
\end{center}
\end{table}

The effect of  $ \left[ Ca^{2+} \right] $ on the beating pattern
can also be studied qualitatively. As mentioned in the
introduction, $ \left[ Ca^{2+} \right] $ has a strong influence on
the beating pattern. Calcium concentration variations are at the
basis of the shock responses of many organisms, changing the
ciliary-type beating into a flagellar-type beating in
\textit{Chlamydomonas}, or switching the directions of effective
stroke and recovery stroke in \textit{Paramecium}, or  in
reversing the direction of the "wave" propagation on the flagellum
and thus reversing the direction of the movement in
\textit{Chritidia} \cite{Sugrue}.

This last example can be explained qualitatively within our
approach. In \textit{Chritidia}, both directions are possible for
the deformation wave propagation. Calcium may affect the chosen
mode of beating, allowing the system to choose $ \Omega_{c}^{(1)}
$ and its tip-to-base pattern instead of $ \Omega_{c}^{(3)} $.

On the other hand, calcium is also likely to change the
attachment/ detachment rate (and thus change the parameter $ \nu
$) or the boundary conditions at the base of the cilium  (and thus
change $ k $). In Chlamydomonas, calcium has a contractile effect
on the striated fibers connecting the basal bodies of the two
flagella \cite{Hayashi}. These changes induce a change in the
beating pattern, and may result in a switch from base-to-tip to
tip-to-base wavelike propagation.

\section{Left-right beating symmetry breaking}
\label{breaking}

\begin{figure}
\begin{center}
\includegraphics[angle=0, width=7cm]{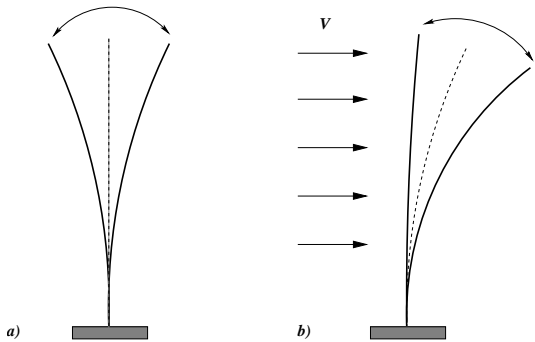}
\end{center}
\caption{\label{Asymmetry} Effect of an external flow $ \vec{V} $
on the beating of a single cilium.  $ a) $ Symmetrical beating.
 $ b) $ broken symmetry due to the external flow.}
\end{figure}

In the presence of a transverse external flow, the beating can no
longer be symmetrical as sketched on Fig. \ref{Asymmetry}. The
cilium tends to beat faster and quite straight in the direction of
the flow, whereas it comes back slower and more curved against the
flow. This looks like a two-phases beating with an effective and a
recovery stroke.

If the beating is asymmetrical, the cilium exerts a force in the
fluid that can itself produce a flow. In a certain range of
parameters, one can therefore expect  that a continuous flow is
spontaneously generated by hydrodynamic interactions between
cilia: an assembly of cilia, beating symmetrically, is able to
break spontaneously this left-right symmetry of the beating to
create a global flow. This idea of a spontaneous breaking
spontaneously of the left-right symmetry has already been
suggested in \cite{Marco} with a more abstract system (called
rowers) having two internal energy states.

In this section,  we first study the effect of an external
velocity imposed by the experimentalist on the beating symmetry of
a single cilium. We then consider an array of aligned cilia and
determine the conditions under which this assembly of cilia breaks
its left-right symmetry and generates a global flow. Metachronal
coordination between cilia naturally emerges from hydrodynamic
couplings as a local minimum of the oscillation threshold
$\Omega_{c}$.

\subsection{External breaking of the beating symmetry: cilium submitted to an external flow }

We impose an external flow $ \vec{V}=V\vec{e_{x}}$ along the $ x $
axis for simplicity. It is found experimentally that the velocity
above the cilia sub-layer is time independent and uniform
\cite{Liron}, justifying our choice. This flow is in this first
part externally fixed and we consider the limit of vanishingly
small flows.

The force per unit length exerted by the cilium on the fluid $
\vec{f}_v [\vec{X}(s)] $  depends on the external velocity  $
\vec{V} $.
\begin{equation}
\label{force1}
\vec{f}_v [\vec{X}(s)]=(\xi_{\parallel}\vec{t}\ \vec{t}+
\xi_{\perp}\vec{n}\ \vec{n})\ (\partial_{t}\vec{X}(s)-\vec{V})
\end{equation}
The equation of motion \ref{flagelle} reads then:
\begin{equation}
\label{Equation}
\partial_{t}\vec{X}=\vec{V}+\frac{\vec{n}}{\xi_{\perp}}(-\kappa \dddot {\psi} - a \dot f +
\dot \psi \tau)
+ \frac{\vec{t}}{\xi_{\parallel}}(\kappa \dot \psi \ddot \psi + a
\dot \psi f + \dot \tau)
\end{equation}
The boundary conditions are the same as  in the absence of the
neighboring cilia and are given by Eq. \ref{limit1}, \ref{limit2},
\ref{basalsliding}.

Following the same procedure  as for a cilium in the absence of
flow, we find the equation of motion for the deformation of the
cilium $h$:
\begin{equation}
\label{flagelle4} \xi_{\perp} \partial_{t}h =\xi_{\perp}V -\kappa
\ddddot h - a \dot f -  \xi_{\parallel}V(h \ddot h +
\dfrac{\xi_{\perp} }{2 \xi_{\parallel}} \dot h ^{2}) + \mathcal{O}
(\vert h \vert^{3}  )
\end{equation}
The introduction of the external flow breaks the $ h
\longrightarrow -h $ symmetry (or left-right symmetry) introducing
in \ref{flagelle1} terms of zeroth  and second order in $h$ in the
equation of motion. The boundary conditions do not depend on the
external flow.

As above, we expand the deformation of the cilium $h$ in Fourier
components in time. Using the same notations as before, the
equation of motion of the Fourier components can be written as
\begin{equation}
\label{flagelle5} \ddddot {\bar{h}}_{n}+\bar\chi_{n}\
\ddot{\bar{h}}_{n} +in\bar \omega\ {\bar{h}}_{n}=\bar V
\delta_{0,n}-\dfrac{\bar V}{2}(\bar \xi \  {\bar{h}}
\ddot{\bar{h}} +{\dot{\bar{h}}}^{2} )_{n}
\end{equation}
for $ n=0,1,2 $ and where we have introduced the new dimensionless
parameters:
$$ \bar V=\dfrac{\xi_{\perp}L^{3}}{\kappa}V \ \ \ \ \ \ \ \ \ \
\bar \xi = \dfrac{2 \xi_{\parallel}}{\xi_{\perp}} $$
In the limit of small external velocities,  we have neglected
terms of order $ \bar V^{2} $.

The equation for the first mode is identical to Eq.
\ref{flagelle31}, with the same boundary conditions. At this order
in $ \bar V $, the fundamental mode is not affected by the
external flow. Consequently, the oscillation threshold and the
beating frequency are the same as in the absence of flow and the
Fourier component $h_1$ is given by Eq. \ref{h}.

\begin{figure}
\begin{center}
\includegraphics[ width=7cm]{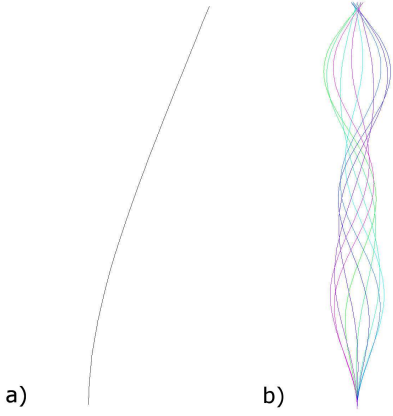}
\end{center}
\caption{\label{h0h2} $ a)$ Average position of a cilium which is
curved in the direction of the flow,
$\bar{h}_{0}(\bar{s})=<\bar{h}(\bar{s},t)>$. $ b)$ Second Fourier
component of the deformation $ 2
\Re[\bar{h}_{2}(\bar{s})e^{2i\omega t}]$ at different times during
a beating period. The scale is dilated: $ \vert \bar{h}_{2} \vert
\ll 0.1 $ with the parameters  $ \bar{\xi}=1 $, $
\mathcal{A}_{1}=1/70 $ and $ \bar V = 1 $.}
\end{figure}

The zeroth Fourier component $\bar{h}_0$ gives the average
deformation of the cilium. It is a solution of
\begin{equation}
\label{flagelle50} \ddddot{\bar{h}}_{0}-\bar K\ \ddot{\bar{h}}_{0}=\bar V -
\dfrac{\bar V}{2}\left[  \bar \xi \  ( {\bar{h}}_{1} \ddot{\bar{h}}_{1}^{*}+
{\bar{h}}_{1}^{*} \ddot{\bar{h}}_{1}) + \vert \dot{\bar{h}}_{1} \vert ^{2}\right]
\end{equation}
with the same boundary condition as before. Nevertheless,
$\bar{h}_0$ does not vanish at first order in velocity because of
the broken symmetry due to the external flow which is reflected in
the right hand side of Eq. \ref{flagelle50}. The complete solution
for $ {\bar{h}}_{0} $ is rather tedious to obtain and lengthy. We
do not display it here explicitly. We write it as the sum of two
contributions; $ {\bar{h}}_{0}^{Eq} $, corresponds to the
curvature of the cilium under the flow $ V $ at equilibrium, i.e.
in the absence beating ($ {\bar{h}}_{1}=0 $), and
${\bar{h}}_{0}^{ATP} $, corresponds to the corrections to this
equilibrium deformation due to the beating when there is enough
ATP in the medium $\bar{h}_{0}=
{\bar{h}}_{0}^{Eq}+{\bar{h}}_{0}^{ATP}$. If as above, we ignore
the elasticity of the nexins ($ \bar K \rightarrow 0 $) :
\begin{eqnarray}
\nonumber {\bar{h}}_{0}^{Eq}(\bar s)&=&\dfrac{\bar V}{24}\bar s ^{2}
(\bar s ^{2}- 4 \bar s +6) \\
{\bar{h}}_{0}^{ATP}(\bar s)&=&\dfrac{\bar V}{2}
\mathcal{A}_{1}^{2} \phi_{0} (\bar s)
\end{eqnarray}
where $\mathcal{A}_1 $ is the amplitude of the first Fourier mode
of the oscillation defined in Eq. \ref{h} and $\phi_{0}(\bar s) $
is a linear combination of exponentials. In the limit $ V=0 $, $
{\bar{h}}_{0}=0 $ as expected. The average deformation of the
cilium is plotted on Fig. \ref{h0h2} which shows the bent shape
under the action of the external flow.

The second Fourier component gives the asymmetry of the beating.
It is obtained from the equation of motion
\begin{equation}
\label{flagelle52} \ddddot{\bar{h}}_{2}+\bar\chi_{2}\
\ddot{\bar{h}}_{2}+2i\bar \omega\ {\bar{h}}_{2}=- \dfrac{\bar
V}{2}\left[  \bar \xi  \ {\bar{h}}_{1} \ddot{\bar{h}}_{1} +
 {\dot{\bar{h}}_{1}}^{2}\right]
\end{equation}

We do not give here the lengthy explicit expression of
$\bar{h}_{2}$ but we write it as
\begin{equation}
\nonumber {\bar{h}}_{2}(\bar s)=\dfrac{\bar V}{2} \mathcal{A}_{1}^{2} \phi _{2}(\bar s)
\end{equation}
where $ \phi_{2}(\bar s) $ is a linear combination of
exponentials. Here also, in the limit  $ V=0 $, $ {\bar{h}}_{2}=0
$. The plot $ {\bar{h}}_{2} $ against $ \bar s $ at different
times on Fig. \ref{h0h2}, leads to a complicated pattern.

\begin{figure}
\begin{center}
\includegraphics[angle=90, width=2.5cm]{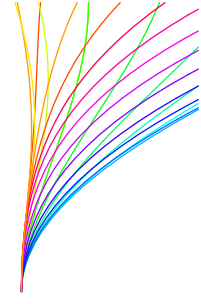}
\end{center}
\caption{\label{hzoom} Beating pattern at the basis of the cilium
($ \bar{s} \in [0,0.2] $) with the parameters $ \bar{\xi}=1 $, $
\mathcal{A}_{1}=1/5 $ and $ \bar V = 2 $ : the cilium beats faster
in the direction of the flow and slower in the opposite direction
around a curved average position.}
\end{figure}

The total deformation of the cilium $ {\bar{h}}(\bar s,t)\simeq
{\bar{h}}_{0}(\bar s) + {\bar{h}}_{1}(\bar s)e^{i\omega_{c}t}+
{\bar{h}}_{2}(\bar s)e^{2i\omega_{c}t} + c.c.$ is plotted against
$ \bar s  $ at different times equally spaced on Fig. \ref{hzoom}.
In order to stress the fact that the beating is easier and faster
in the direction of the flow, and more difficult and slower
against the flow, we have chosen rather large values of the
parameters, $\bar{\xi}=1 $, $ \mathcal{A}_{1}=1/5 $ and $ \bar V =
2 $, and we plot $ {\bar{h}}(\bar s,t) $ for  $ \bar s \in [0,0.2]
$ on Fig. \ref{hzoom}.

The external flow thus breaks the left-right symmetry in two ways.
First the average position of the cilium is not the vertical axis
 but a cilium curved in the direction of the flow. Second,
the beating itself is no longer left-right symmetric: the cilium
goes faster in the direction of the flow and comes back slower
against the flow. The beating pattern looks like a two-phases
beating with an effective stroke and a recovery stroke. The
external flow may therefore be an important factor in the
asymmetry of the beating.

Another important result, is that, because the beating propagates
a base-to-tip deformation, the curved cilium exerts a finite
average force in the fluid in the direction of the flow. Thus, if
an external flow breaks the left-right beating symmetry, the cilia
create a force in its direction and can amplify this flow. This is
the basis of the left-right spontaneous symmetry breaking that we
discuss in the next section.

The external flow is not always the only source of symmetry
breaking. If it were so, a \emph{Paramecium} would always go in
the same direction once it started moving. This is not the case,
this organism is able to go backward when it bumps into an
obstacle thanks to the release of calcium that reverses the
beating.

The calculations of this section have been made with a velocity
$\vec{V}$ uniform over the cilium length. This is not consistent
with the presence of a cell wall where the cilium is anchored.
Nevertheless, the main idea was to study how an external flow can
break the beating symmetry in the simplest way. Similar
calculations can be performed with a linearly varying velocity
$\vec{V}= U \bar{z}$; they do not lead to any new physical
effects.

\subsection{Spontaneous breaking of the beating symmetry: array of aligned cilia}

We now consider a regular array of cilia on a cell body, beating
all in the same direction. Starting from a symmetrical beating, we
show that the left-right symmetry is spontaneously broken within a
certain range of the parameters controlling the beating due to the
hydrodynamic couplings between cilia.

\subsubsection{Equations of motion}

For a cilium located in the $ xy $ plane at position $ \vec{r} $,
we call $ \vec{V}[\vec{X}(s)] $ the velocity created by the other
cilia at the point $\vec{X}(s) $ of arclength $s$. The equation of
motion of the cilium is similar to that obtained previously with
an external flow field and we write up to third order in $h$ as
\begin{equation}
\label{harray}
\xi_{\perp} \partial_{t} h=-\kappa \ddddot h - a \dot f +
\xi_{\perp} \vec{n}.\vec{V} + \mathcal{O}(h^{3},\vec{t}.\vec{V}h)
\end{equation}
where the projection of the local external velocity on the cilium
normal is
\begin {equation}
\vec{n}.\vec{V}=V_x(1-\dot h^2/2-V_z\dot h+ \mathcal{O}(h^{4}))
\end{equation}
The boundary conditions for the motion are the same as in the
previous section.

The velocity $ \vec{v}_{j}[\vec{X}_{i}(s_{i})] $ created at
arclength $ s_{i} $ of the cilium $ i $ by a cilium $ j $ is given
by
\begin{equation}
\nonumber \vec{v}_{j}[\vec{X}_{i}(s_{i})]=\int_{0}^{L}ds_{j} \
\textbf{G}[\vec{X}_{i}(s_{i}),\vec{X}_{j}(s_{j})].\vec{f}_{j}[\vec{X}_{j}(s_{j})]
\end{equation}
where $ \textbf{G} $ is the second order hydrodynamic tensor given
by Eq. \ref{Gtensor} and $ \vec{f}_{j}=\vec{f}_{j}^{beat} $ is the
force per unit of length created by the beating of the cilium $ j
$. The total velocity at the arclength $ s_{i} $ of the cilium $ i
$ is thus given by
\begin{equation}
\vec{V}[\vec{X}_{i}(s_{i})]= \sum_{j\neq i} \vec{v}_{j}[\vec{X}_{i}(s_{i})]=
\vec{V}(\vec{r}_{i},s_{i},t)
\end{equation}

As in section \ref{rotation}, we consider the limit $ L \ll d $
and we only keep terms of the second order in $ s/r $, $ r $ being
the distance between cilia so that
\begin{equation}
\nonumber
\textbf{G}.\vec{f}=\left( [\textbf{G}.\vec{f}_{x}].\vec{e}_{x}\right)  \vec{e}_{x}+
\mathcal{O}(s^{3}/r^{3})
\end{equation}
This means that only the velocity along the $ x $ axis created by
the component $ f_{x} $ of $ \vec{f} $ plays a role and that we
can ignore the other component $V_z$ of the velocity. Using the
notations of Fig. \ref{Lattice}, and noting that $ z=s+
\mathcal{O}(h^{2})$, we obtain
\begin{equation}
\label{vitesse}
 V_{x}(\vec{r}_{i},s_{i},t)=\dfrac{3s_{i}}{2\pi
\eta}\sum_{j\neq i} \dfrac{\cos ^{2} \theta _{ij}}{\vert
\vec{r}_{i}-\vec{r}_{j} \vert ^{3}} \int_{0}^{L} ds_{j}
{f_{j}}_{x} (s_{j}) s_{j}+\mathcal{O}(h^{3},\frac{s^{3}}{r^{3}})
\end{equation}

As in the previous sections, we expand the velocity, the force and
the cilium deformation in Fourier modes in time. For simplicity,
we only consider here the first two Fourier components and do not
look at the Fourier component $h_2$ that characterizes the
asymmetry of the beating. The Fourier components of the velocity
are related to the Fourier components of the force by
\begin{equation}
\nonumber V_{n}(s_{i})\simeq \dfrac{3s_{i}}{2\pi \eta}\sum_{j\neq
i}  \dfrac{\cos ^{2} \theta _{ij}}{\vert \vec{r}_{i}-\vec{r}_{j}
\vert ^{3}}\ \int_{0}^{L} ds_{j} {f_{j}}_{n} (s_{j}) s_{j}
\end{equation}

The Fourier components of the force  $ f_{0}=<f_{x}> $ and $ f_{1}
$ are calculated using the expression of $ f^{beat} $ and its
average over one time period given by Eq. \ref{fxbeataverage} in
the small movements approximation:
\begin{eqnarray}
\nonumber f_{0}&\simeq & 2\omega (\xi_{\perp}-\xi_{\parallel})
\Im
[2\dot{h}_{0}h_{1}\dot{h}_{1}^{*}-\dot{h}_{1}^{*}\int_{0}^{s}du
\dot{h}_{1}(u)\dot{h}_{0}(u)]
\\
\label{fzero}
f_{1}&\simeq & i\omega \xi_{\bot} h_{1}
\end{eqnarray}
where $\Im$ is the imaginary part of a complex number.

We assume that all cilia are identical, and that they all beat
with the same pattern. The only difference in the beating patterns
of cilia $ j $ and $ i $ is a possible phase difference that we
call $ \varphi_{ij} $. Defining
\begin{equation}
\nonumber I_{n}=\int_{0}^{L} ds_{i} f_{n} (s_{i}) s_{i}
\end{equation}
and dropping the index $ i $, we write the Fourier components of
the velocity as
\begin{eqnarray}
\label{ vzero} V_{0}(s)&=&\dfrac{3I_{0}s}{2\pi\eta} \sum_{j\neq i}
\dfrac{\cos^{2} \theta_{ij}}{\vert \vec{r}_{i}-\vec{r}_{j} \vert
^{3}}=\dfrac{3\mathcal{K}I_{0}}{2\pi\eta d^{3}}\ s
\\
\nonumber V_{1}(s)&=&\dfrac{3I_{1}s}{2\pi\eta} \sum_{j\neq i}
\dfrac{\cos^{2} \theta_{ij}}{\vert \vec{r}_{i}-\vec{r}_{j} \vert
^{3}}\ e^{i\varphi_{ij}}=\dfrac{3\mathcal{K}[\left\lbrace
\varphi_{ij}\right\rbrace ]I_{1}}{2\pi\eta d^{3}}\ s
\end{eqnarray}
The geometrical constant  $ \mathcal{K} $ is given by Eq.
\ref{constanteK} for a square lattice of cilia spaced by $ d $.
The constant $\mathcal{K}[\left\lbrace \varphi_{ij}\right\rbrace
]$ depends on the relative phases between the cilia. If  the
phases $ \varphi_{ij} $ are randomly distributed, then $
\mathcal{K}[\left\lbrace \varphi_{ij}\right\rbrace] \simeq 0 $ and
$ V_{1} =0 $. There is no oscillating component of the velocity.

On the contrary, because we know that metachronism occurs in an
array of beating cilia, we choose a constant phase difference $
\varphi $ between two consecutive cilia in the direction of the
plane of beating : $\varphi_{i,j+1}-\varphi_{i,j}=\varphi $. This
is the case for simplectic and antiplectic metachronal
coordination. We only consider those cases (and not laeoplectic or
dexioplectic metachronism) here. Experimentally, for
\textit{Opalina} (simplectic) and \textit{Pleurobrachia}
(antiplectic) that both have planar beatings, no metachronal wave
in the transverse direction of the beating can be seen
\cite{Murase}.

\begin{figure}
\begin{center}
\includegraphics[angle=0, width=8cm]{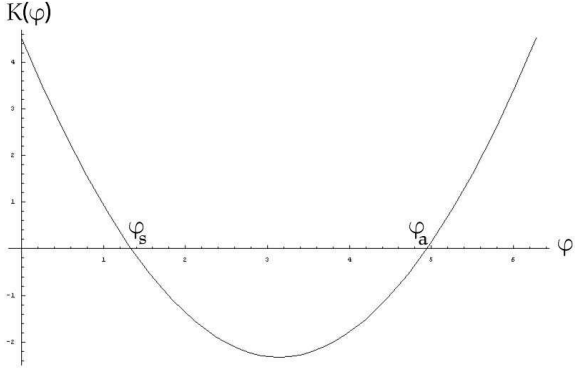}
\end{center}
\caption{\label{K(phi)} $ \mathcal{K}(\varphi) $ over one period
($ \varphi\in[0,2\pi] $). Some remarkable values:
$\mathcal{K}(0)=\mathcal{K}_{max}\simeq 4.52 $;
$\mathcal{K}(\pi)=\mathcal{K}_{min}\simeq -2.32 $;
$\mathcal{K}(\varphi_{s})=\mathcal{K}(\varphi_{a})=0$ with
$\varphi_{s}\simeq 1.34$ and $\varphi_{a}\simeq 4.94$.}
\end{figure}

We stress that we do not impose the phase difference $ \varphi $.
The system is free to adjust its phase. We then write $
\mathcal{K}[\left\lbrace
\varphi_{ij}\right\rbrace]=\mathcal{K}(\varphi) $ with
\begin{equation}
\mathcal{K}(\varphi)=\sum_{(k,l)\neq (0,0)}\dfrac{k^{2}e^{ik\varphi}}{(k^{2}+l^{2})^{5/2}}
\end{equation}
Note that $ \mathcal{K}(0)=\mathcal{K} $. The function $
\mathcal{K}(\varphi) $ is plotted on Fig. \ref{K(phi)} for a
lattice of $ 10^{6} $ cilia.

Note that for two particular values of $ \varphi $ that we denote
by
 $ \varphi_{s} $ and $ \varphi_{a} $ this function vanishes, $
\mathcal{K}(\varphi_{s})=\mathcal{K}(\varphi_{a})=0 $, as in the
case where the relative phases of the cilia are randomly
distributed. This  corresponds to a constant flow with no
oscillating component.

We now define the two dimensionless velocities $ U $ and $
u(\varphi) $ by
\begin{equation}
\label{Uu}
\bar U = \dfrac{\xi_{\perp}L^{3}}{\kappa} \dfrac{3\mathcal{K}I_{0}L}{2\pi\eta d^{3}}\ \ \ \ \ \ \
\ \ \bar{u}(\varphi)=
\dfrac{\xi_{\perp}L^{3}}{\kappa}\dfrac{3\mathcal{K}(\varphi)I_{1}L}{2\pi\eta d^{3}}
\end{equation}
The equations of motions of the Fourier components $\bar{h}_{0}$
and $ \bar{h}_{1}$ can then be written as
\begin{eqnarray}
\label{ordrezero}
\nonumber \ddddot{\bar{h}_{0}}-\bar{K}\ddot{\bar{h}}_{0} &=& \bar{U}\bar{s} \\
\ddddot{\bar{h}_{1}}+\bar{\chi}_{1} \ddot{\bar{h}}_{1}+i\bar{\omega}\bar{h}_{1} &=&
\bar{u}(\varphi)\bar{s}
 \end{eqnarray}

In writing Eq. \ref{ordrezero}, we only kept the term $
\bar{U}\bar{s}= \mathcal{O}(h^{3}) $ that breaks the left-right
symmetry and that lead to $ h_{0}\neq 0 $, ignoring any other term
that would not create a macroscopic motion.

\subsubsection{Beating pattern and metachronal waves}

We first study the equation of motion of the first Fourier mode in
Eq. \ref{ordrezero}, which corresponds to the oscillatory motion
of the cilium. The right hand side of this equation of motion does
not vanish due to the existence of an oscillatory external flow
due to the other cilia. Note however that we have not treated in
details the hydrodynamic interactions for one cilium and that we
have only taken them into account through the two local friction
coefficients $ \xi_{\perp} $ and $ \xi_{\parallel} $. We are here
more interested in the qualitative aspects of the coordination
between cilia than in the accurate calculation of the flows
created by each cilium.

The general solution of Eq. \ref{ordrezero} can be written as  $
\bar{h}_{1}=\bar{h}_{1}^{h}+\bar{h}_{1}^{p}$ with
\begin{eqnarray}
 \nonumber
  \bar{h}_{1}^{h}(\bar{s}) &=& A_{1}e^{q_{1}\bar s}+B_{1} e^{-q_{1}\bar s}
+ C_{1}e^{q_{2}\bar s} + D_{1}e^{-q_{2}\bar s}
\\
  \bar{h}_{1}^{p}(\bar{s})  &=& \dfrac{\bar{u}(\varphi)}{i\bar{\omega}}\bar{s}
\end{eqnarray}
It is convenient to rewrite the external velocity as $
\bar{u}(\varphi)=i\bar{\omega}\mathbb{C}_{1}\gamma(\varphi)$ with
\begin{equation}
\gamma(\varphi)=\dfrac{3\mathcal{K}(\varphi)\xi_{\perp}L^{3}}{2\pi\eta
d^{3}}\ \ \ \ \ \ \ \mathbb{C}_{1}=\int_{0}^{1}d\bar{s}\
\bar{h}_{1}(\bar{s})\bar{s}
\end{equation}
The constant $ \mathbb{C}_{1} $ can be determined self-consistently
as it varies linearly with  $\bar{h}_{1} $. We obtain
\begin{equation}
\bar{h}_{1}(\bar{s})=\sum_{i}A_{i}(e^{q_{i}\bar{s}}+\beta(q_{i},\varphi)\bar{s})
\end{equation}
with
\begin{equation}
\beta(q_{i},\varphi)=\dfrac{\gamma(\varphi)}{1-\gamma(\varphi)/3}
\dfrac{q_{i}e^{q_{i}}-e^{q_{i}}+1}{q_{i}^{2}}
\end{equation}
The effect of the hydrodynamic interactions between cilia is
embodied here in the coefficient $\gamma(\varphi) $. The variation
of this coefficient with the phase difference $\varphi$ is similar
to that of  $ \mathcal{K} (\varphi)$. The limit where $
\gamma(\varphi)=0 $, leads back to the previous situation were one
cilium is beating alone;  it may however correspond to the finite
phase shifts between cilia $ \varphi=\varphi_{s}$ or $\varphi_{a}
$.

\begin{figure}
\begin{center}
\includegraphics[angle=0, width=8cm]{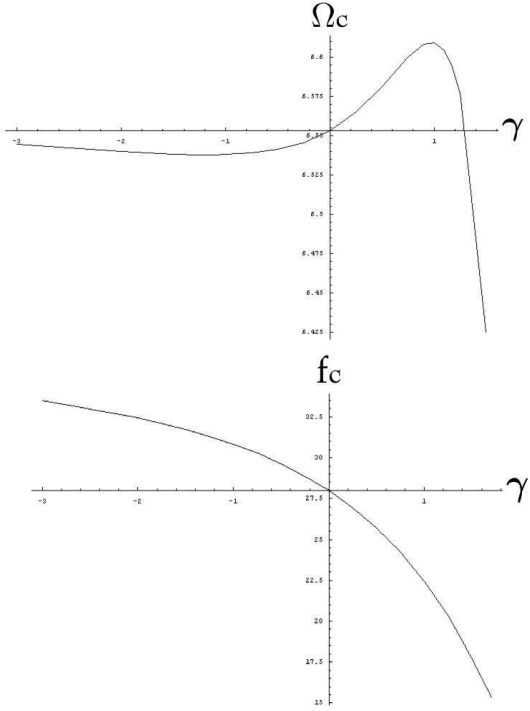}
\end{center}
\caption{\label{Omegagamma} Oscillation threshold $ \Omega_{c} $
and critical frequency $ f_{c} $ as functions of $ \gamma $ for $
\gamma \in [-3,2] $. $ \Omega_{c} $ has a local minimum that
corresponds to the existence of metachronal waves.}
\end{figure}

The four boundary conditions on $ \bar{h}_{1} $ can as before be
written in a matrix form and the oscillation threshold and the
beating  frequency can be determined as the zeros of a determinant
insuring the consistency of this matrix equation. This leaves an
unknown amplitude of the beating motion that could only be
calculated by expanding the equation of motion to higher order.
The beating pattern can then be written as
\begin{equation}
\nonumber
\bar{h}_{1}(\bar{s})=\mathcal{A}_{1}[\mathcal{E}_{\varphi}(q_{1},\bar{s})
+ b_{1}\mathcal{E}_{\varphi}(-q_{1},\bar{s})
+ c_{1}\mathcal{E}_{\varphi}(q_{2},\bar{s})
+ d_{1}\mathcal{E}_{\varphi}(-q_{2},\bar{s})]
\end{equation}
with
\begin{equation}
\nonumber
\mathcal{E}_{\varphi}(q,\bar{s})=e^{q \bar{s}}+\beta(q,\varphi)\bar{s}
\end{equation}
The values of both the oscillation threshold $ \Omega_{c}$ and the
frequency $\bar {\omega}_{c}$ depend on the phase shift between
cilia $ \varphi $, through $ \gamma(\varphi) $. We first discuss
the variation of this bifurcation point with the constant
$\gamma(\varphi)$
 which is a more convenient variable.
On Fig. \ref{Omegagamma}, we plot $ \Omega_{c} $ and the critical
frequency $ f_{c} $ against $ \gamma $.

There is a local minimum of $ \Omega_{c} $ for $ \gamma^* \simeq
-1.15 $ and a local maximum for $ \gamma \simeq 1$. The beating
frequency $ f_{c} $, is a decreasing function of $ \gamma $.

We here need a selection criterion that determines the value of
the phase shift between cilia. The simplest conjecture for the
selection criterion is that the system chooses the local minimum
of $ \Omega_{c} $ corresponding to $ \gamma^{*} \simeq -1.15 $.
This corresponds to a metachronal wave propagating in the assembly
of cilia, as widely confirmed by experimental observations
(\cite{cells.de, Machemer} for instance).

With this selection criterion, the oscillation threshold is
$\Omega_{c}\simeq 6.538\  10^{-9}$ and the critical frequency is $
f_{c}\simeq 31 Hz $. The hydrodynamical couplings between cilia
decrease the oscillation threshold $ \Omega_{c} $ and increase the
critical frequency $ f_{c} $.  The coordination between cilia
favors cilium beating  by creating a metachronal wave
corresponding to $ \gamma < 0 $.

\begin{figure}
\begin{center}
\includegraphics[angle=90, width=3cm]{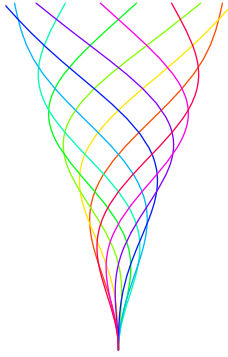}
\end{center}
\caption{\label{h1meta} Beating pattern of a cilium in an array in
the presence of a metachronal wave. The pattern is different from
that of an isolated cilium mostly around the basis. The first
Fourier component $ 2\Re[\bar{h}_{1}e^{i \omega t}] $ at various
time steps during a period is plotted. The parameters are $
\mathcal{A}_{1}=1/70 $; the maximum deformation is $ \bar{h}_{max}
\simeq 0.15 $.}
\end{figure}

The beating pattern is slightly changed as shown on Fig.
\ref{h1meta} where we have plotted $ 2\Re[\bar{h}_{1}e^{i \omega
t}] $ at different time steps with the same amplitude $
\mathcal{A}_{1}=1/70 $.

The phase difference $ \varphi^{*} $ between two consecutive cilia
corresponding to $ \gamma^{*} \simeq -1.15 $ depends on the values
of the parameters. If we take $ \eta=4\eta_{w} $ and $ d/L=1 $ so
that our calculations remain consistent and in order to be close
to what is observed experimentally, then $\mathcal{K}(\varphi^{*})
\simeq -0.07$ which yields $ \varphi^{*}\simeq \pm 1.37 \simeq \pm
0.44 \pi$. This value corresponds to a wavelength $ \lambda = 4.6
d \sim 5 d$ for the metachronal waves or approximatively $6$
cilia, which is the correct order of magnitude ( the wave length
is $ 7 $ cilia in \cite{Machemer}).

\subsubsection{Global flow and left-right symmetry breaking}

We now discuss the left-right symmetry breaking and the appearance
of a global flow. We solve Eq. \ref{ordrezero} for the zeroth
Fourier component of the deformation, with the same boundary
conditions as before, in the limit $ \bar{K}\rightarrow 0 $. We
obtain
\begin{equation}
\bar{h}_{0}(\bar{s})=\bar{U}\dfrac{\bar{s}^{2}}{6}
(1-\dfrac{\bar{s}}{2}+\dfrac{\bar{s}^{3}}{20})
\end{equation}
The cilium oscillates around a curved average position $
\bar{h}_{0}\neq 0 $ if $ \bar{U} \neq 0 $, if there exists a
global flow. We show below that this is possible within a certain
range of parameters.

We define the two dimensionless functions
\begin{eqnarray}
\nonumber H_{0}(\bar s)=\dfrac{\bar{h}_{0}(\bar s)}{\bar U}&=&
\dfrac{\bar{s}^{2}}{6}(1-\dfrac{\bar{s}}{2}+\dfrac{\bar{s}^{3}}{20})
\\
\nonumber H_{1}(\bar s)=\dfrac{\bar{h}_{1}(\bar
s)}{\mathcal{A}_{1}}&=&\mathcal{E}_{\varphi}(q_{1},\bar{s}) +
b_{1}\mathcal{E}_{\varphi}(-q_{1},\bar{s})
\\
\nonumber  &+& c_{1}\mathcal{E}_{\varphi}(q_{2},\bar{s}) +
d_{1}\mathcal{E}_{\varphi}(-q_{2},\bar{s})
\end{eqnarray}
The determination of the average velocity $U$ requires the
calculation of the integral  of $ I_{0} $ defined in Eq.
\ref{fzero}; we obtain
\begin{equation}
I_{0}=2\mathbb{C}_{\varphi}{\mathcal{A}_{1}}^{2}
\bar{U}(\xi_{\perp}-\xi_{\parallel})L^{3}\omega
\end{equation}
with
\begin{equation}
\label{cphi}
\mathbb{C}_{\varphi}=\int_{0}^{1}d\bar{s}\ \Im[2
\dot{H}_{0}H_{1}\dot{H}^{*}_{1}-\dot{H}^{*}_{1}\int_{0}^{\bar{s}}
d\bar{s}'\dot{H}_{0}\dot{H}_{1}]
\end{equation}
which can be numerically calculated knowing $ {H}_{0} $
and $ {H}_{1} $. $ \mathbb{C}_{\varphi} $ depends on $ \varphi $
through $ {H}_{1} $. Using the value of $\phi$ corresponding
to metachronal waves, we obtain
\begin{equation}
\nonumber \mathbb{C}_{\varphi}\simeq 34.5
\end{equation}

A self-consistent equation is then be obtained for the average
velocity $ \bar U$
\begin{equation}
\label{selfU}
\bar U=\dfrac{3 \mathcal{K}
\mathbb{C}_{\varphi}(\xi_{\perp}-\xi_{\parallel}) L^{3}}{\pi \eta d^{3}}
{\mathcal{A}_{1}}^{2} \bar{\omega}\  \bar U
\end{equation}
If $ \mathbb{C}_{\varphi}<0 $ this equation has the only solution
$ U=0 $ and no global flow can exist, the left-right symmetry is
not broken.
If $ \mathbb{C}_{\varphi}>0 $ this equation can have two extra non
zero solutions $ U \neq 0 $ corresponding to a global flow along
the $x$ axis given by
\begin{equation}
\nonumber
<V(\bar{s},t)>=V_{0}(\bar{s})= U \bar{s}
\end{equation}
and the left-right symmetry is then broken.

The condition for appearance of a global flow is
\begin{equation}
\label{leputaindecritere}
 \dfrac{3\mathcal{K}\mathbb{C}_{\varphi}{\mathcal{A}_{1}}^{2}}{\pi}
 \dfrac{\xi_{\perp}-\xi_{\parallel}}{\eta}
 \dfrac{\xi_{\perp}L^{4}\omega}{\kappa}\dfrac{L^{3}}{d^{3}}>1
\end{equation}
As for the oscillation amplitude, our calculation only gives the
threshold of appearance of the global flow. A determination of the
actual value of the velocity would require an expansion of the
equations of motion to higher orders.

\section{Discussion and concluding remarks}

We have studied in this paper  how  hydrodynamic interactions
between cilia contribute to the coordination of the beating motion
in ciliated cells. Three major effects have been studied, the
spontaneous alignment of an array of cilia, the breaking of the
symmetry of the beating and the appearance of a macroscopic flow
and the existence of metachronal waves. We have shown for all
these problems that there exist a dynamic transition where
symmetry is broken and where a coordination between the beating of
neighboring cilia appears.

Our work is based on several simplifying approximations that we
believe make the analysis tractable analytically but that should
preserve the essential physical effects. We only studied
hydrodynamic interactions between distant cilia that can be
treated by introducing simple distribution of forces in the fluid
to describe the motion of one cilium. This is rarely true
experimentally but the hydrodynamic interactions between closer
cilia are even stronger and strongly favor the transitions that we
study. We have replaced the complex architecture of the axoneme by
two microtubules sliding against one another under the action of
dynein motors which are described by a two state model for
molecular motors as done earlier by Camalet and Jülicher. This is
a rather sketchy description but it allows a calculation of the
internal forces that drive the cilium motion and it gives some
physical insight. Future work will have to take into account the
nine-fold symmetry of the axoneme and the influence of its central
doublet. Finally, we have only considered small amplitude beating.
This is sufficient to determine the oscillation threshold but it
does not allow a quantitative comparison between the  calculated
beating and the experimental one that often occur far from any
threshold. All our results are qualitatively consistent with the
experimental observations and for example the beating frequency is
close to both the experimental ones and to the ones obtained in
numerical simulations \cite{Gueron}.

The essential result of our work is the natural emergence of
metachronal waves and of a macroscopic flow created by an array of
cilia if the amplitude of beating is large enough. The criterion
for appearance of the global uniform component of the flow given
by Eq. \ref{leputaindecritere} requires only very small amplitudes
($\mathcal A_1 \geq 5\ 10^{-4}$) which means that as long as the
left-right symmetry is broken, a macroscopic flow should appear.
An essential ingredient for the macroscopic flow to appear is that
the constant $ \mathbb{C}_{\varphi} $ defined in Eq. \ref{cphi} be
positive so that the average force created by one cilium favors
the flow and does not oppose it (which occurs if $ \varphi=0$).

As long as we allow a constant phase shift between neighboring
cilia we observe metachronal coordination as a consequence of
hydrodynamic interactions and of the internal beating mechanism of
the cilium. A selection criterion is then needed for these waves.
We have conjectured that the existing metachronal wave is the one
that corresponds to the local minimum of the oscillation
threshold. A more complete calculation that goes far beyond the
scope of this work would have to consider the nucleation of the
metachronal wave and to determine the fastest growing wave. One of
the interesting predictions of our calculation is that the
existence of metachronal waves leads to a flow which is far more
stationary than if all the cilia were beating in synchrony. The
oscillating component of the flow is proportional to the constant
$\mathcal{K}(\varphi) $ (see Eq. \ref{Uu})  which has a much
smaller value when metachronal waves exist
($\mathcal{K}(\varphi^{*})\simeq -0.07 $) than if all cilia are
beating in synchrony   ($ \mathcal{K}(0)=\mathcal{K}\simeq 4.52
$). Metachronism thus contributes to the creation of a very steady
movement of swimming organisms that could for example make easier
the detection of the organism environment.

Our most important conclusion is the idea that metachronism and
the existence of macroscopic flow around ciliated organisms can
exist as self-organized phenomena driven by hydrodynamic
couplings.  We must stress however that other mechanisms could be
at the origin of these cooperative effects.

\textbf{Aknowledgements }: We thank A. Hilfinger, P.
Dupuis-Williams, N. Spassky, M. Cosentino Lagomarsino, J. Prost
and M. Bornens for useful discussions.

\section*{Appendix I: Average force created by a single beating cilium in a viscous fluid}
\label{appendixI}

The aim of this appendix is to calculate the force and momentum
averaged over one time period created by a general periodic
beating of a single cilium. We make two assumptions: the beating
is planar and there is a stationary  external  flow. In section
\ref{breaking}, the average flow is created by the neighboring
cilia.

\begin{figure}
\begin{center}
\includegraphics[angle=0, width=8cm]{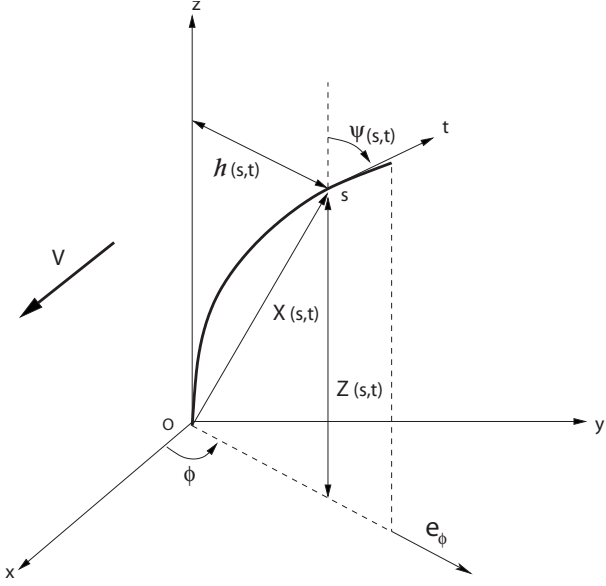}
\end{center}
\caption{\label{Notations} Sketch of a beating cilium in a plane
at an angle $ \phi $ with the direction of the external flow $ V
$.}
\end{figure}

We call $ \phi $ the angle between the plane of beating and the
direction of the external flow $ \vec{V} $ that we take along the
$ x $ axis. The cilium of length $L$ is located at the origin and
it is fixed at its basis. We denote by  $ h(s,t) $ the distance
between a point at arclength $ s $ on the cilium and the $ z $
axis at time $ t $ and by $ Z(s,t) $ the distance between a point
at the arclength $ s $ on the cilium and the $ xy $ plane. The
angle between the tangent vector  $ \vec{t} $ to the cilium and
the $ z $ axis is denoted by $\psi (s,t) $ (see Fig.
\ref{Notations}). The coordinates of the tangent vector are $
\vec{t}=(\cos \phi \sin \psi, \sin \phi \sin \psi, \cos \psi)  $.
The angle $ \psi $ is related to the cilium deformation $ h $ by
$\sin \psi = \partial_{s} h$

The point on the cilium at the arclength $ s $
is located at position  $ \vec{X}=(x,y,z) $, with:
\begin{eqnarray}
\nonumber x &=&\cos \phi \int_{0}^{s} du\  \sin \psi (u,t)= h(s,t)\cos \phi  \\
\nonumber y &=&\sin  \phi\int_{0}^{s} du\  \sin \psi (u,t)= h(s,t)\sin \phi  \\
\nonumber z &=&\int_{0}^{s} du\  \cos \psi (u,t)=Z(s,t)
\end{eqnarray}
The velocity of this point is calculated by derivation with
respect to time, $ \vec{v}=\partial_{t} \vec{X} $.

The force per unit length exerted by the cilium on the fluid
expressed in the Frenet basis  $ (\vec{t},\vec{n},\vec{b}) $ is:
\begin{equation}
\vec{f}=(\xi_{\parallel}\vec{t}\ \vec{t}+\xi_{\perp}\vec{n}\ \vec{n}+
\xi_{\perp}\vec{b}\ \vec{b})(\vec{v}-\vec{V})
\end{equation}
where $ \xi_{\parallel} $ and $ \xi_{\perp} $ are the two local
friction coefficients for tangential and normal motion
respectively. We decompose this force as a sum of two forces, $
\vec{f}^{beat} $ depending on the local velocity and $
\vec{f}^{flow} $ depending on the external flow velocity and
calculate the average force over a beating period
 $<\vec{f}>\ =\dfrac{1}{T}\int_{0}^{T} dt\ \vec{f}(t)$.

The average beating force $ <\vec{f}^{beat}> $ can be explicitly
calculated
\begin{eqnarray}
\label{fxbeataverage}\nonumber <f_{x}^{beat}>&=&(\xi_{\perp}-
\xi_{\parallel})\dfrac{\cos \phi}{2}\int_{0}^{s} du<\partial_{t}
\psi (u) \cos \Delta(u,s) >
\\
\nonumber <f_{y}^{beat}>&=&(\xi_{\perp}-\xi_{\parallel})
\dfrac{\sin\phi}{2}\int_{0}^{s} du<\partial_{t} \psi (u) \cos
\Delta(u,s)>
\\
\nonumber <f_{z}^{beat}>&=&-(\xi_{\perp}-\xi_{\parallel})
\dfrac{1}{2} \int_{0}^{s} du<\partial_{t} \psi (u) \sin
\Delta(u,s)>
\\
\Delta(u,s)&=& 2\psi (s)-\psi (u)
\end{eqnarray}
This force is proportional to $ (\xi_{\perp}-\xi_{\parallel}) $ as
mentioned in section \ref{rotation}. The difference between the
two local friction coefficients $ \xi_{\perp} $ and $
\xi_{\parallel} $ is at the basis of the flow generation by an
assembly of beating cilia. Indeed, this is because the shape of
the beating in the effective stroke is different from that in the
recovery stroke that a force can be exerted in the fluid on
average.

The average force due to the external flow is
\begin{eqnarray}
\nonumber <f_{x}^{flow}>&=&(\xi_{\perp}-\xi_{\parallel}) V
\cos^{2} \phi <\sin^{2}\psi >-\xi_{\perp} V
\\
\nonumber <f_{y}^{flow}>&=&(\xi_{\perp}-\xi_{\parallel}) V \cos
\phi \sin \phi <\sin^{2}\psi >
\\
\nonumber <f_{z}^{flow}>&=&(\xi_{\perp}-\xi_{\parallel}) V <\sin
\psi  \cos \psi  >
\end{eqnarray}
It is important to note that $<f_{x}^{flow}(s)><0 $ : this force
opposes the flow. The last term of $ f_{x}^{flow} $ is a static
term, whereas the first positive term depends on the beating
pattern  and reduces the effects of this static term. In an
assembly of cilia, the external velocity is due to the beatings of
the other cilia which are themselves created by the forces on
these cilia.

In section \ref{rotation}, we introduce a viscous coefficient $
\alpha $ which characterizes the tendency for a cilium, beating in
a plane at an angle $ \phi $ with the flow, to align with the
other cilia. A torque along the $ z $ axis due to the flow $
M_{z}^{flow}=-\alpha U \sin \phi $ is exerted on this cilium. We
now express $ \alpha $ as a function of the cilium beating
pattern. We call $ m_{z}=-(\vec{X} \times \vec{f}).\vec{e}_{z} $
the torque along $ z $ exerted by the fluid on the cilium per unit
of length (the minus sign is due to the fact that $ \vec{f} $ is
the force exerted by the cilium on the fluid). The local torque
per unit length exerted by the fluid on the cilium is
\begin{equation}
\nonumber m_{z}(s,t)=-\xi_{\perp}V h(s,t) \sin \phi
=-\xi_{\perp}L \bar{h}(s,t) \bar{Z}(s,t) U \sin \phi
\end{equation}
where we have used the dimensionless coordinates $\bar{s}=s/L$,
$\bar{h}=h/L$. The total momentum along $ z $ averaged over time,
is obtained by integration
\begin{equation}
M_{z}=-\xi_{\perp}L^{2} \int_{0}^{1}
d\bar{s}<\bar{h}(\bar{s},t)\bar{Z}(\bar{s},t)> \ U \sin \phi
\end{equation}
This defines the friction coefficient $ \alpha $:
\begin{equation}
\label{alpha} \alpha=\xi_{\perp}L^{2}
\int_{0}^{1}d\bar{s}<\bar{h}(\bar{s},t)\bar{Z}(\bar{s},t)>
\end{equation}
which can be calculated if the motion of the cilium is known.

\section*{Appendix II}
In this appendix, we derive the equations satisfied by the Fourier
components of the deformation $h$ of a single beating cilium and
we determine the threshold of spontaneous oscillations of the
cilium.

\subsubsection*{Fourier mode expansion}

Axoneme beating is  periodic and can be studied by expansion in
Fourier modes  in time of all the physical parameters:
\begin{equation}
\nonumber h(s,t)=\sum_{n=-\infty}^{\infty}  h_{n}(s) e^{i n\omega t}
\end{equation}
The definition is similar for the other parameters. Starting from
Eq. \ref{shearforce} and Eq. \ref{basalsliding}, we obtain the
Fourier components:
\begin{eqnarray}
\nonumber f_{n}(s)&=&-(K +in\omega \lambda) \Delta_{n}-\frac{1}{l}
\int_{0}^{l}d\xi\ P_{n}(\xi)\partial_{\xi}W_{1}
\\
\label{delta} \Delta_{0n}&=&\dfrac{1}{k + i n\omega \gamma}
\int_{0}^{L} d s  f_{n}(s)
\end{eqnarray}

In order to determine the non linear relationship between $ f $
and $ \Delta $, we follow the lines of \cite{Prost2} and write:
\begin{equation}
\nonumber f_{n}= f_{n}^{(0)} + \sum_{l}
f_{nl}^{(1)}\Delta_{l}+\sum_{lm}
f_{nlm}^{(2)}\Delta_{l}\Delta_{m}+\mathcal{O} (\Delta^{3})
\end{equation}
The coefficients $ f_{n,n_{1},...,n_{k}}^{(k)} $ can be calculated
by first rewriting Eq. \ref{Proba} as
\begin{equation}
\label{Proba2} P_{n}=R \delta_{n,0}-\dfrac{i \omega}{\nu(1+
\Omega)}\sum_{lm}l\delta_{n,l+m}\Delta_{l} \partial_{\xi}P_{m}
\end{equation}
where
\begin{equation}
\nonumber R=\rho \dfrac{\omega_{2}(\xi)}{\omega_{1}+
\omega_{2}}=\rho \dfrac{1+\Omega \sin^{2}(\pi \xi/l)}{1+\Omega }
\end{equation}
is the static probability $ (\omega=0) $, corresponding to a
medium with not enough ATP to generate the beating. Inserting the
ansatz
\begin{equation}
\nonumber P_{n}=R \delta_{n,0}+\sum_{l} P_{nl}^{(1)}\Delta_{l}+
\sum_{lm} P_{nlm}^{(2)}\Delta_{l}\Delta_{m}+\mathcal{O}
(\Delta^{3})
\end{equation}
into Eq. \ref{Proba2}, we obtain a recursion relation for the $
P_{n,n_{1},...,n_{k}}^{(k)} $:
\begin{equation}
\nonumber P_{n,n_{1},...,n_{k}}^{(k)}=-\dfrac{i
\omega}{\nu(1+\Omega )}\sum_{m}n_{k}\delta_{n,n_{k}+m}
\partial_{\xi}P_{m,n_{1},...,n_{k-1}}^{(k-1)}
\end{equation}
that now allows us to calculate $ f_{n,n_{1},...,n_{k}}^{(k)} $.

Our choice of a symmetric potential $ W_{1} $ imposes that a
change $ \Delta \rightarrow -\Delta $ must change  $ f \rightarrow
-f $. This symmetry imposes thus $ f^{(2k)}=0$. The only
non-vanishing coefficient at linear order is $ f_{nl}^{(1)}=
\chi(\Omega, n \omega)\delta_{n,l}$ with
\begin{eqnarray}
\label{chi1} \chi(\Omega,\omega)=
-K-\lambda i \omega + \dfrac{\pi^{2} \rho U}{2 l^{2}}
\dfrac{i \Omega  \omega}{(1+\Omega)((1+\Omega)\nu+i\omega)}
\end{eqnarray}
The force and the sliding displacement are thus related by
\begin{equation}
\label{slidingh} f_{n}=\chi(\Omega, n \omega) \Delta_{n}+
\mathcal{O}(\vert\Delta\vert^{3})=
 \chi(\Omega, n \omega)( \Delta_{0n}+a\dot
h_{n}+ \mathcal{O}(\vert h\vert ^{3}))
\end{equation}
This relationship \ref{slidingh} models the response of the
molecular motors to the bending of the axoneme.

From Eq. \ref{delta} we obtain
\begin{equation}
\label{deltah} \Delta_{0n}=\dfrac{\chi(\Omega, n \omega)a}{k
+in\omega-\chi(\Omega, n \omega)L}h_{n}(L)+\mathcal{O}(\vert
h\vert ^{3})
\end{equation}
We solve the equation of motion of the cilium (Eq.\ref{flagelle1})
for each order of the Fourier expansion.

\subsubsection*{Equation of motion of the Fourier modes}

We look for an approximate solution of the form
\begin{equation}
\nonumber h(s,t)\simeq h_{0}(s)+h_{1}(s)e^{i\omega t}+h_{2}(s)
e^{2i\omega t}+ c.c.
\end{equation}
At linear order, there is no coupling between the modes and using Eq.\ref{flagelle1},
the equation of motion of the $n^{th}$ Fourier
component reads
\begin{equation}
\label{flagelle2} \ddddot h_{n}+\dfrac{\chi(\Omega, n
\omega)a^{2}}{\kappa}\ddot h_{n}+ i\dfrac{n\omega
\xi_{\perp}}{\kappa}h_{n}=0
\end{equation}
It is convenient to introduce the dimensionless variables:
\begin{equation}
\nonumber \bar s=s/L \ \ \ \ \ \ \ \ \bar
\omega=\dfrac{\xi_{\perp}L^{4} }{\kappa}\omega \ \ \ \ \ \ \ \ \ \
\bar \chi _{n}=\bar \chi(\Omega, n \bar \omega) =
\dfrac{a^{2}L^{2}}{\kappa}\chi(\Omega, n \omega)
\end{equation}
In dimensionless form, Eq. \ref{chi1} can be written
\begin{equation}
\nonumber
\bar \chi(\Omega, \bar \omega)=-\bar K-\bar \lambda i
\bar \omega +  \dfrac{\pi^{2}}{2}\bar \rho \bar U \dfrac{i\
\Omega \bar \omega}{\bar \nu + i \bar
\omega}
\end{equation}
with
\begin{equation}
\nonumber \bar K=\dfrac{a^{2}L^{2}}{\kappa} K\ \ \ \ \ \ \ \bar
\lambda = \dfrac{a^{2}}{\xi_{\perp}L^{2}}\lambda\ \ \ \ \ \ \bar
\nu=\dfrac{\xi_{\perp}L^{4} }{\kappa}\nu\ \ \ \ \ \ \ \ \bar
U=\dfrac{a^{2}L^{2}}{\kappa l^{3}}U
\end{equation}
We have anticipated here the fact that  $ \Omega \ll 1 $.

Defining, $\bar{h}=h/L$, and denoting by a dot the derivation with
respect to $\bar s$, we obtain the equation of motion  Eq.
\ref{flagelle3} and the boundary conditions given by Eq.
\ref{bound0},\ref{bound1}.

\subsubsection*{Beating motion}

In the absence of external flow only the first Fourier component of $h$ does
not vanish and satisfies the equation of motion :
\begin{equation}
\label{flagelle31} \ddddot{\bar{h}}_{1}+\bar\chi_{1}\
\ddot{\bar{h}}_{1}+i\bar \omega\ {\bar{h}}_{1}=0
\end{equation}
where the relevant dimensionless parameters are
\begin{eqnarray}
\nonumber \bar\chi_{1}=\bar\chi(\Omega,\bar \omega)\ \ \ \ \ \ \ \
\ \ \ \ \bar \Gamma_{1}=\dfrac{\bar\chi_{1} ^{2}}{\bar k-\bar
\chi_{1}+i\bar\gamma \bar \omega }
\end{eqnarray}
The boundary conditions are given by Eq. \ref {bound0} and
\ref{bound1} for $ n=1 $. The solution to this linear equation is
a superposition of exponentials given by Eq. \ref{h}. The four
boundary conditions on $ {\bar{h}}_{1} $ can be written in a
matrix form:
\begin{equation}
\label{coeff}
\mathbf M_{1}(\Omega,\bar \omega).\mathbf A_{1}=0
\end{equation}
where $\mathbf A_{1}$ is the vector made by the amplitudes of the
exponentials in Eq. \ref{h} and the matrix $\mathbf M_{1}$ is
given by
\begin{displaymath}
\mathbf M_{1}(\Omega,\bar \omega)=
\left[ \begin{array}{cccc}
1 & 1 & 1 & 1 \\
q_{1} & -q_{1} & q_{2} & -q_{2} \\
\mathcal{F}(q_{1}) & \mathcal{F}(-q_{1}) & \mathcal{F}(q_{2}) &
\mathcal{F}(-q_{2})
\\
q_{1}^{2}e^{q_{1}} & q_{1}^{2}e^{-q_{1}} & q_{2}^{2}e^{q_{2}} &
q_{2}^{2}e^{-q_{2}}
\end{array} \right]
\end{displaymath}
with
\begin{equation}
\nonumber \mathcal{F}(q)=e^{q}(q^{3}+\bar\chi_{1} q+\bar\Gamma_{1})
\end{equation}
The system \ref{coeff} has non trivial solutions only if
\begin{equation}
\label{det}
\det \mathbf M_{1}(\Omega,\bar \omega)=0
\end{equation}
Since Eq. \ref{det} is a complex equation, it determines both the
oscillation threshold $ \Omega_{c}$ and the dimensionless beating
frequency $\bar \omega_{c} $.

\bibliography{Biblionew}

\end{document}